\documentclass[Phys]{SciPost}

\usepackage{graphicx}
\usepackage{amsmath}
\usepackage{amssymb}
\usepackage{yfonts}
\usepackage{slashed} 
\usepackage{hyperref}

\usepackage{marvosym}
\usepackage{wasysym}
\usepackage{xcolor}

\newcommand{\Sn}{N}
\newcommand{\Se}{S}

\newcommand{\ombar}{\bar \omega}

\newcommand{\rhoA}{\rho_E }

\newcommand{\sig}{\sigma}

\newcommand{\bg}{\bar g}
\newcommand{\bh}{h}
\newcommand{\gn}{g}
\newcommand{\rmr}{{\rm \hspace{0.2mm} r}}
\newcommand{\rms}{{\rm s}}

\renewcommand{\mid}{{\rm mid}}
\newcommand{\reg}{{\rm r}}
\newcommand{\varrhoq}{{\varrho_q}}
\newcommand{\fqrho}{f_{\hspace{-0.3mm}\varrho}}

\newcommand{\inh}{{\rm i}}
\renewcommand{\hom}{{\rm h}}

\newcommand{\hnul}{{(0)}}
\newcommand{\heen}{{(1)}}

\newcommand{\Mc}{M_c}
\newcommand{\qc}{q_i}
\newcommand{\qi}{q_i}
\newcommand{\Qc}{Q_c}
\newcommand{\cQc}{{\cal Q}_c}

\newcommand{\mmin}{{\hspace{0.2mm}-\hspace{0.2mm}}}
\newcommand{\pplus}{{\hspace{0.2mm}+\hspace{0.2mm}}}
\newcommand{\plus}{{\hspace{0.4mm}+\hspace{0.4mm}}}
\newcommand{\iss}{{\hspace{0.2mm}=\hspace{0.2mm}}}

\newcommand{\zp}{{\rm zp}}

\newcommand{\refl}{\ref}

\newcommand{\vph}{\varphi}

\renewcommand{\d}{{\rm d}}

\newcommand{\itf}{{\it f}}

\newcommand{\diag}{{\rm diag}}

\newcommand{\rhom}{\rho_\lambda}
\newcommand{ \brhoL}{\bar \rho_\lambda}

\newcommand{\half}{{\frac{1}{2}}}

\renewcommand{\max}{{\rm max}}

\newcommand{\myskip}[1]{}

\newcommand{\vA}{{\bf A}} 
\newcommand{\vB}{{\bf B}} 
\newcommand{\vE}{{\bf E}}

\newcommand{\vJ}{{\bf J}}

\newcommand{\ex}{{\rm ex}}

\newcommand{\p}{\partial}   
\newcommand{\onedot}{\,\,\,}   
\newcommand{\twodots}{\,\,\,\,}   
   
\newcommand{\ed}{\onedot}
\newcommand{\ednu}{{\onedot\nu}}

\newcommand{\td}{\twodots}

\newcommand{\Om}{\Omega}
\newcommand{\BEQ}{\begin{eqnarray}}   
\newcommand{\EEQ}{\end{eqnarray}}   
\newcommand{\BEA}{\begin{eqnarray}}   
\newcommand{\EEA}{\end{eqnarray}}   
\newcommand{\nn}{\nonumber }   
\renewcommand{\d}{{\rm d}}

\newcommand{\vth}{\vartheta}   
   
\newcommand{\eps}{\varepsilon}   
\newcommand{\om}{\omega}

\newcommand{\mn}{{\mu\nu}}

\renewcommand{\diag}{{\rm diag}}

\newcommand{\cC}{{\cal C}}

\newcommand{\cE}{{\cal E}}

\newcommand{\cK}{{\cal K}}

\newcommand{\cO}{{\cal O}}

\newcommand{\cQ}{{\cal Q}}

\newcommand{\sth}{s_\theta}

\newcommand{\Gam}{\Gamma}

\begin{document}

\begin{center}
{\Large \textbf{Exact solutions for black holes with a smooth quantum core} }
\end{center}

\begin{center}
Theodorus Maria Nieuwenhuizen
\end{center}

\begin{center}
{Institute for Theoretical Physics,  University of Amsterdam 
 \\ Science Park 904, 1090 GL  Amsterdam, The Netherlands} \\   
* t.m.nieuwenhuizen@uva.nl
\end{center}

\begin{center}
version \today
\end{center}

\myskip{
} 


\section*{Abstract}
A class of exact solutions are presented for the interior of black holes of solar mass and beyond. 
In a core enclosed by the inner horizon, the binding energy released by dissolution of the pre-collapse nuclei
is stored in electrostatic and zero point energy. Gravitational collapse is prevented by their negative pressures.
In the mantle, the region between the inner and event horizons, there is a standard vacuum.
Accounting for the rest masses of the up and down quarks and electrons leads to corrections at the per cent level.
Spherically symmetric fluctuations  have a spectrum without unstable modes.
A surface layer with a charge current can be present on the outer side of the inner horizon; a layer of opposite charges 
on the event horizon can make an extremely charged black hole neutral.
Merging of an extremal black hole with another extremal one or with a neutron star may produce electromagnetic fireworks.

\vspace{10pt}
\noindent\rule{\textwidth}{1pt}
\tableofcontents\thispagestyle{fancy}
\noindent\rule{\textwidth}{1pt}
\vspace{10pt}

\hfill{Excellit in rebus explicandis quae intelligi non possunt\footnote{He excels in explaining subjects that can not be understood}}

\vspace{3mm}

\section{Introduction}

Black holes (BHs) are fascinating objects in the cosmos and in theories. Astrophysical BHs originate from 
gravitational collapse of stars. Their existence is motivated by direct observation \cite{penrose2020nobel,genzel2021forty} 
and detection of  gravitational waves generated in merging events \cite{weiss2019ligo,barish2019ligo,thorne2019ligo}.
Their mass may range from a few solar ones to billions of them, championed by
Ton 618 with its 66 billion solar masses \cite{shemmer2004near}.

It is standard to describe the BH  interior by a metric in empty
space \cite{schwarzschild1916gravitationsfeld,reissner1916eigengravitation,weyl1917gravitationstheorie,
nordstrom1918energy,jeffery1921field,Kerr1963gravitational,newman1965note}. 
They contain a singularity that cannot be described in general relativity, but will be supposedly resolved by a not yet known theory of quantum gravity.
In the description of gravitation as a field in Minkowski space-time and Euclidean space \cite{landau1975classical,babak1999energy,nieuwenhuizen2007einstein},  
all mass is concentrated in the singularity.
 That our present theories would not be able to describe these otherwise classical objects 
is felt as unsatisfactory by many. It motivated  Bardeen \cite{bardeen1968non} to study a BH with a smooth interior metric,  
supported by matter with a regular equation of state.  Various follow up studies were carried out 
 \cite{dymnikova1992vacuum,hayward2006formation,nieuwenhuizen2008exact,nieuwenhuizen2010bose,
frolov2014information,casadio2015thermal,mazur2015surface,simpson2020regular},
with an exact solution in nonlinear electrodynamics  \cite{ayon2000bardeen}.

Further progress was recently made when we proposed that the interior is described by standard model physics \cite{nieuwenhuizen2021interior}.
As a start, it was assumed that in the stellar explosion, more electrons are ejected, so that the remaining core collapses to
a positively charged BH. For large central density, the precollapse protons, neutrons and nuclei will dissolve into quarks, 
thereby releasing their $\sim 99\%$ binding energy. 
It will produce a Higgs condensate and thermal particles spontaneously produced by pair creation in the standard model.

Here we show that in the non-rotating case, the regularized Schwarzschild metric allows a class of
exact solutions when no thermal particles are present, so that their temperature vanishes.
An essential role will be played by the electromagnetic field and the {\it tuneable zero point energy}, a possibility not considered before.
They are nontrivial in a {\it core} bounded by the {\it inner horizon}, which occurs in models with a smooth center.
 We employ the term {\it mantle} for the region between the inner horizon and event horizon.
It is an empty space-time, described by a standard metric, in our charged, non-rotating case, by the Reissner-Nordstr\"om metric.
The situation is sketched in figure 1.
We also consider an extra surface layer at the inner horizon $R_i$, and likewise at the event horizon $R_e$.

The property of Schwarzschild BHs that all mass goes to the singularity means in our case only that it goes to the  smooth core.
 While time and radius reverse their role inside the Schwarzschild BH,  their role is normal inside the inner horizon, that is, in the core,
a property already present in the Reissner-Nordstr\"om BH.
Since time and particle orbits are normal (just as in ordinary Newtonian motion)  in the core, there is no a priori reason to form a singularity.

The setup of this work is as follows. In section \ref{estimates} we consider physical estimates for the setup
and in section \ref{secEineqsEMT} the mathematical formulation.
In section \ref{objections} we address some objections against the approach.
In sections \ref{electro} and \ref{zeZPE} we present physical explanations for the ingredients of the setup.
Approximate exact solutions are discussed in section \refl{exactsols} and a full numerical solution in section \ref{sec-col-mat}.
 Fluctuations and their stability is investigated in section \ref{sec-stab}.
Boundary layers at the inner and event horizons are discussed in section \ref{boundaries}.
We close with a summary, discussion and outlook.

 \renewcommand{\thesection}{\arabic{section}}
\section{Physical estimates}
\setcounter{equation}{0}
\renewcommand{\theequation}{2.\arabic{equation}}
\renewcommand{\thesection}{\arabic{section}}

\label{estimates}

Assuming a central energy density $\lesssim\lambda v^4$, where $\lambda=0.129$ is the Higgs self coupling 
and $v=246$ GeV the vacuum expectation value of the Higgs field, leads for a volume $R^3\sim (GM)^3$ 
to a mass estimate $M\gtrsim M_\ast=m_P^2/\lambda^{1/2}v^2$, with $m_P=1/\sqrt{G}$ the Planck mass in natural units
$\hbar=c=1$. 
This combination of fundamental constants comes out 
as $M_\ast=1.5 $ Neptune masses or 25 Earth masses,  opening the door for describing astrophysical BHs
of solar mass and beyond.

 In a supernova explosion electrons are more easily ejected than protons, therefore we assume that the matter for the BH is positively charged.
In an imploding stellar core of mass $M$ and charge $Q$, with profiles $M(r)$ and $Q(r)$ and uniform charge--to--mass ratio,
the ratio of forces on a chunk ($ch$) of matter, $[Q(r)Q_{\it ch}/r^2]/[GM(r) M_{\it ch}/r^2]=(Qm_P/M)^2$, 
allows BH formation just up to the extremal charge $Q_\max =M/m_P$. So
this ``back of the envelope'' estimate  offers wiggle room for the onset of charged, few solar mass BH-core formation.

The charge mismatch is small. In the extremal case, it reads: $Q/eN_e=M/em_PN_e=m_N/em_P=10^{-18}$.
This tiny fraction is still important in BHs since the Coulomb force is so much stronger that the Newton force.

Sauter-Schwinger creation of an electron-positron pair \cite{sauter1931verhalten,schwinger1951gauge}
 in an electric field $E$ is possible for $E(\hbar/m_ec)\gtrsim m_ec^2$.
 For a BH with charge $Q=qQ_\max$, the electric field at  the event horizon, $E\sim Q/R^2 \sim (qM/m_P)/ (GM)^2=qm_P^3/M$,
allows pair creation for $M\lesssim qm_P^3/m_e^2=6.2\,10^{6}qM_\odot$, whereupon the electrons condense on the BH and the positrons escape to infinity.
 Equivalently, this leaves for a BH a charge-to-mass ratio $q\lesssim {\rm min}(1,\,1.6\,10^{-7}M/M_\odot)$,
which can be shielded by electrons from accretion.

The Coulomb energy of a charge $e$ at the event horizon of a charged BH,  estimated as 
$eQ/R\sim e(qM/m_P)/(GM)=\sqrt{\alpha}\,qm_P=1.0\,10^{18}q$ GeV, 
is so large that it offers hope for quantum tunnelling effects in BH merging events.

\renewcommand{\thesection}{\arabic{section}}
\section{The metric and the stress energy tensor}
\setcounter{equation}{0}
\renewcommand{\theequation}{\thesection\arabic{equation}}
\renewcommand{\thesection}{\arabic{section}.}

\label{secEineqsEMT}

 We express the line element $\d s^2 =g_\mn \d r^\mu \d r^\nu$  in coordinates $r^\mu=(t,r,\theta,\phi)$ with $\mu=0,1,2,3$, as
  \cite{nieuwenhuizen2021interior}
\BEQ\label{dssqTN}
\d s^2 \iss \mmin \Sn ^2\bar \Se  \d t^2\pplus \frac{1}{\bar \Se}\d r^2\mmin r^2(\d\theta^2\pplus\sth^2\d\phi^2),\hspace{1mm} 
\qquad \sth\equiv \sin\theta ,
\EEQ
 where $N=N(r)$, $S=S(r)$ and $\bar \Se  =\Se  -1$.
The Schwarzschild metric has $\Sn =1$, $\Se  =2GM/r$ and
the Reissner Nordstr\"om (RN) metric  $\Sn =1$, $\Se  =2GM/r-GQ^2/r^2$.
The latter has an event ($e$) and inner ($i$) horizon located at
\BEQ \label{ReRi=}
R_{e}=GM+G\sqrt{M^2-\cQ^2},\quad 
R_{i}=GM-G\sqrt{M^2-\cQ^2},\quad \cQ\equiv m_PQ . 
\EEQ
For general $N,S$ the Einstein tensor $G^\mu_\ednu$  is diagonal with $G^2_{\ed2}=G^3_{\ed3}$ due to spherical symmetry. 

We express the stress energy tensor in the 3 parameters $\rhom $, $\rhoA $ and $\sigma_\vth$, 
 \BEQ \label{TMnrhomA}
&& \hspace{-4mm} 
T^\mu_\ednu = \frac{G^\mu_\ednu}{8\pi G} = \brhoL \delta^\mu_\ednu+\rhoA  \cC^\mu_\ednu
+\sigma_\vth (U^\mu U_\nu -\frac{\delta^\mu_\ednu}{4}),
 \quad  \cC^\mu_\ednu =\diag(1,1,\mmin 1,\mmin 1) ,
 \nn \\&& \hspace{-4mm} 
\brhoL=\rhom+\frac{1}{4}\tau_\vth, \qquad \sigma_\vth=\rho_\vth+p_\vth , \qquad \tau_\vth=\rho_\vth-3p_\vth , 
\EEQ
where $U^\mu$ is the velocity vector of thermal matter with energy density $\rho_\vth$, isotropic pressure 
$p_\vth$ and $T^\mu_{\vth\,\nu}=(\rho_\vth+p_\vth)U^\mu U_\nu-p_\vth\delta^\mu_\ednu$.
In the core it holds that $U^\mu=\delta^\mu_0/\Sn \sqrt{1-\Se  }$.
Given the functions $N,S$, our task is now to provide a physical meaning for the parameters
\BEQ \label{EMT=}
\label{rhoL=}
&& \hspace{-6mm}
\brhoL  =\frac{2\Se  \pplus 4r \Se  '\pplus r ^2\Se  ''}{32\pi G r^2} \pplus 
\frac{\Sn ' }{\Sn }\frac{4 \bar \Se  \pplus 3 r \Se  '}{32\pi G r} \pplus \frac{\Sn ''}{\Sn }\frac{\bar \Se  }{16\pi G} , 
\\ &&  \hspace{-6mm}
\label{rhoA=}
\rhoA =\frac{2 \Se  -r^2\Se  ''}{32 \pi  G r^2 }+\frac{\Sn ' }{\Sn }\frac{2 \bar \Se   -3 r \Se  ' }{32 \pi  G r} -\frac{\Sn ''}{\Sn } \frac{\bar \Se  }{16 \pi  G }  ,
\\ &&  \hspace{-6mm}
\label{sigth=}
\sigma_\vth=-\frac{\Sn '\bar \Se  }{4\pi Gr\Sn } =\frac{\Sn '(1-\Se  )}{4\pi Gr\Sn } . 
\EEQ
As standard in GR, their $1/\lambda^2$ scaling upon setting $r\to\lambda r$ implies that for any solution with mass $M$, 
there is a family of solutions with masses $\lambda M$.

 For suitable functions $N(r),S(r)$, we will seek a physical explanation for the quantities
$\rho_\lambda$, $\rhoA $ and $\sigma_\vth$  inside  a core enclosed by the inner horizon $R_i$,
while this core is surrounded by an empty mantle
ranging up to the event horizon $R_e$, see figure 1. Surface layers at $R_{i,e}$ will also be considered. 

\begin{figure}\label{figrhozpRiRe} 
\centerline{ \includegraphics[width=3cm]{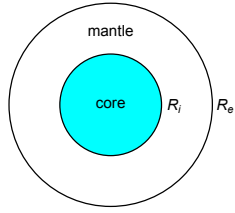}}
\caption{
A core with zero point energy and net charge from  standard model particles is enclosed by the inner horizon $R_i$. 
The core is surrounded by an empty mantle up to the event horizon at $R_e$. }
\end{figure}

\renewcommand{\thesection}{\arabic{section}}
\section{Standard objections}
\setcounter{equation}{0}
\renewcommand{\theequation}{\thesection.\arabic{equation}}

\label{objections}

Before considering solutions of the above equations, we discuss how some standard objections against our BH models with hair
(i. e., various parameters describing the interior) can be circumvented.

\subsection{The Kretschmann curvature invariant}

The Schwarzschild singularity at $r=0$ is best understood from gauge invariant quantities, that is to say, quantities which are equal 
in all coordinate systems,  i. e., for all observers. 
The Kretschmann curvature invariant  $\cK=R^{\kappa\lambda\mu\nu}R_{\kappa\lambda\mu\nu}$ takes the value
\BEQ \label{curvinvar}
\cK=4\frac{\Se  ^2}{r^4} + 2 \frac{\Se  '{}^2}{r^2} + 
 2\frac{ (\Sn \Se  '+2 \Sn ' \bar \Se    )^2}{  r^2 \Sn ^2} + 
 \frac{( \Sn  \Se  ''+3 \Sn '\Se  ' +  2\Sn ''  \bar \Se   )^2}{\Sn ^2} .
   \EEQ
$\cK$ is regular at the event horizon $r=R_e$, so no physical singularity occurs there, only an apparent one (a horizon) in certain coordinate systems.
But  for $r\to0$ in the Schwarzschild case  $\Sn =1$, $\Se  (r)=2GM/r$, $\cK$ diverges as $1/r^6$, and in the RN case even as $1/r^{8}$,  
which points at a physical singularity at the origin. 
Hence $\cK$ is often considered as the nail in the coffin for an unlimited applicability of GR to macroscopic systems. 

The singularity not being resolved has led to the general opinion that the BH singularity can only be described by quantum gravity.
However, this is based  on uniqueness theorems related to these standard metrics, which disregard the quantum nature of matter.
For our approach based on quantum field theory in curved space time, we assume that the issue as still open.
In doing so, we risk the danger of getting stuck along the way (which did not occur), or of making predictions that are not supported by observations.
But these risks should not prevent the theoretical exploration of the possibility.

Likewise, the Penrose-Hawking theorems on trapping surfaces refer to classical physics,
motivated by their violation by Hawking radiation. We leave aside the question whether they apply 
to our quantum field theoretic description of BHs, and aim to predict effects that can be tested in BH merging events.

In the limit $r\to0$, the curvature invariant (\ref{curvinvar}) diverges when $M\to M(r) \sim r^k$  with $k<3$, while it vanishes when $k> 3$.
We shall stick to the intermediate case $k=3$, where the energy density $M'(r)/4\pi r^2$ remains finite at $r=0$, implying $\Se  \sim r^2$.
With it, $\cK$ remains finite at $r=0$,  so that no physical singularity is expected anywhere.

\subsection{Collapse into the singularity?}

Standard lore is that an observer inside the BH will reach the singularity in finite proper time;
in the Schwarzschild and RN metrics this pertains to the origin. 
Let us therefore consider the free fall equations of neutral particles 
\BEQ \label{ddrt=}
\ddot r^\mu+\Gam ^\mu_{\ed\nu\rho}\dot r^\nu\dot r^\rho=\Gam^0_{\ed\nu\rho}\dot r^\mu\dot r^\nu\dot r^\rho,
\qquad \dot r^\mu=\frac{\d r^\mu}{\d t},
\EEQ
which arise from ${\d^2r^\mu}/{\d\tau^2}+\Gam^\mu_{\ed\nu\rho}({\d r^\nu/}{\d \tau})({\d r^\rho}/{\d \tau})=0$.

For a spherical shell at radius $r(t)$ the $\mu=1$ case has  the integral of motion 
\BEQ
\frac{\cE^2}{m^2}
=\left\vert \frac{\Sn ^4\bar \Se  ^3}{\dot r^2-\Sn ^2\bar \Se  ^2}\right\vert \qquad
\EEQ
so that for a shell of particles falling in from infinity with energy $\cE$ and mass $m$
\BEQ \label{dotrE}
\dot r=-\Sn \vert\bar \Se  \vert\sqrt{1+ m^2\Sn ^2\bar \Se   /\cE^2},
\EEQ
This holds in the exterior and the core  where $-1<\bar \Se  <0$, since $\Se  \ge 0$ in our approach.  
It is consistent to take $\dot r<0$ in the mantle where $\bar \Se  >0$.

The differential proper time for inward motion $\d r<0$,
\BEQ
\d\tau=-\left\vert \frac{g_{00}}{\dot r^2}+g_{11}\right\vert ^{1/2} \d r =-\frac{\Sn \d r}{\sqrt{1+\Sn ^2(\Se  -1)m^2/\cE^2}} ,
\EEQ
has the formal integral $\tau(r)-\tau(r_1)$, with $\tau(r)=\int^r\d\tau$.
For $\cE=m$, $\Sn =1$ and $\Se  \sim r^2$, $\tau(r)\sim-\log r$ is divergent for $r\to0$, but this is a special case.
For $\Sn (0)=1-\sigma$, $\cE=m(1+\eps)$ the  integral $\tau \sim-\log(\sigma+\eps)$ is finite at $r=0$.
So for a spherical shell it takes a finite proper time to reach the origin.

In this metric, there need not be a singularity at the origin.
For a neutral particle in the $z=0$ ($\theta=\half\pi$) plane one also inspects the $\mu=3$ component of (\ref{ddrt=}).
The angular momentum is conserved,
\BEQ \label{Lpart=}
\frac{r^2\dot\phi}{\Sn ^2(1- \Se  ) }=L,
\EEQ
 while energy conservation now takes the form
\BEQ \label{Epart=}
\hspace{-3mm}
\dot r^2+ \Sn ^4(1-\Se  )^3\frac{ L^2}{r^2}+V=0,\hspace{1mm} V(r)\equiv 
\Sn ^2(1-\Se  )^2\left[ \frac{m^2}{\cE^2}\Sn ^2(1-\Se  )-1 \right],
\EEQ
\BEQ \label{Epart=}
\hspace{-3mm}
\dot r^2=\Sn ^2(1-\Se  )^2v_r^2,\qquad
v_r^2=1-\left(\frac{L^2}{r^2}+ \frac{m^2}{\cE^2}\right)\Sn ^2(1-\Se  )
\EEQ
which generalizes (\ref{dotrE}) to finite $L$. In the core and exterior, where time is normal, one has $\vert v_S\vert <1$. 
In the mantle, the equivalent is $v_t=\d t/\d r=1/v_r<1$, while 
\BEQ
\phi'=\frac{\Sn L}{r^2v_r},
\EEQ
In our models, the metric functions are bounded in the core, viz. $0<\Sn (r)\le 1$ and $0\le \Se  (r)\le 1$ 
with $\Se  \sim r^2$ for $r\to0$, so the origin is not reached when $L\neq 0$.
In the core, classical particles move on orbits with conserved angular momentum in the spherically symmetric, finite potential $V(r)$.
For small $L$, the particles essentially pass through the origin.

Contrary to the Schwarzschild situation, our models have a smooth macroscopic core in which
 time and particle orbits are normal, so that $r$ needs not only decrease along the orbit.
 Even spherical mass shells do not collapse into the origin;  particles with $L\to0$ just pass through the origin.
{\it  While an orbit with increasing $r$ is forbidden in the interior of the Schwarzschild BH, 
this is in our case restricted to the mantle.} As said,  orbits are just normal in the core.
 
Having explained that the buildup of a singularity is a peculiarity of the Schwarzschild metric,
we do not see a compelling reason against exploring a description of smooth BH interiors with standard model physics.

 \subsection{The null, weak and dominant energy conditions}

Many approaches in general relativity lead to some would-be energy momentum tensor.
Generally, energy conditions are meant to present simple conditions that would-be matter
should satisfy on physical grounds.
Here we look into  the various conditions for our case of standard model matter.

The {\it null energy condition} demands that 
\BEQ \label{nullec}
T_\mn k^\mu k^\nu\ge 0 ,
\EEQ
for any future pointing null vector $k_\mu$, viz $k^\mu k_\mu=0$ and $k_\eta>0$.
The form (\refl{TMnrhomA}) obeys it since $k^ik_i=g_{ii}k^i{}^2\le 0$ for each spatial $i$,
\BEQ
T^\mu_\ednu k_\mu k^\nu
=-2\rhoA (k^2k_2+k^3k_3)+\sigma_\vth k_\eta^2\ge 0 .
\EEQ

The {\it weak energy condition} demands (\ref{nullec}) for all time-like $k_\mu$, viz  $k^\mu k_\mu=k^2> 0$.
Our thermal fluid indeed satisfies the weak energy condition,
\BEQ \hspace{-6mm}
(\rho_\vth+p_\vth) (U^\mu k_\mu)^2-p_\vth k^2
\iss \rho_\vth k^\eta k_\eta-p_\vth k^ik_i
\iss \rho_\vth k^2 -(\rho_\vth+p_\vth) k^ik_i>0 ,
\EEQ
where both terms are non-negative.
Also the EM part, which involves a $\rhoA >0$,  satisfies the weak energy condition,
\BEQ \hspace{-6mm}
T^\mu_{E\,\nu}k_\mu k^\nu =\rhoA (k^0k_0+k^1k_1-k^2k_2 -k^3k_3) =\rhoA  k^2-2\rhoA (k^2k_2 +k^3k_3) .
\EEQ
Finally,  $T^\mu_{\Lambda\,\nu}=\rho_\lambda\delta^\mu_\ednu$ involves $T^\mu_{\Lambda\,\nu}k_\mu k^\nu=\rho_\lambda k^2$ 
which is positive, provided $\rho_\lambda>0$, as occurs in our cases.
So the weak energy condition is satisfied for all components in our model.

The {\it dominant energy condition} demands that for every future-pointing ($k^0>0$) causal vector field  $k^\mu$
(either timelike or null), the vector field $ (Tk)^\mu=T^\mu_\ednu k^\nu$ must be a future pointing causal vector.
It holds that
\BEQ 
 (Tk)^\mu&=&\bar\rho_\lambda\,(k^0,k^1,k^2,k^3)+\rhoA \,(k^0,k^1,-k^2,-k^3)\nn\\&+& (\rho_\vth k^0,-p_\vth k^1,-p_\vth k^2,-p_\vth k^3),
\EEQ
which is future directed, $(Tk)^0>0$, and timelike, viz.  $(Tk)^\mu (Tk)_\mu\ge0$, since $\rho_\vth\ge p_\vth$. 
Hence, the dominant energy condition is satisfied.

The {\it strong energy condition} $T^\mu_{\ednu}k_\mu k^\nu- \half T^\mu_{\ed\mu}k^2\ge 0$ for time-like $k$ 
is not satisfied by the $\bar \rho_\lambda$ term, which effectively gets reversed in sign, as usual for this condition.
Therefore, this condition is violated near $r=0$, where $\rho_\lambda$ dominates.

That the null, weak, and dominant energy conditions are satisfied is not much of a surprise, 
but rather expected for standard model matter, the true matter in Nature,
on which these conditions are modelled. But it was seen that the local cosmological constant should be positive, 
as it is in our models.

\renewcommand{\thesection}{\arabic{section}}
\section{Electrostatics}
\setcounter{equation}{0}
\renewcommand{\theequation}{\thesection.\arabic{equation}}

\label{electro}

Let  us go back to section \ref{secEineqsEMT} and start with proposing a cause for $\rhoA $. In the core it can arise from a distribution of static charges.
For the potential $A_\mu=\delta^0_\mu A_0(r)$ and $F_\mn=\p_\mu A_\nu-\p_\nu A_\nu$, 
the nontrivial $\mu=0$ component of the Maxwell equation $F^{\nu\mu}_{\td\ed;\nu}=\mu_0J^\mu$ reads   in terms of the field $E=-A_0'/\Sn $ and source $J^\mu=\delta^\mu_0\rho_q/\Sn $  
as $(E'+2E/r)/\Sn =\mu_0 \rho_q/\Sn $. For $\mu_0=4\pi$ the solution is $E(r)=Q(r)/r^2$ 
with enclosed charge $Q(r)=4\pi\int_0^r\d r\, r^2\rho_q(r)$. Setting 
\BEQ
\rho_q(r)=\frac{Q_c}{4\pi R_i^3}f_q(x),\qquad x\equiv \frac{r}{R_i}\le 1,
\EEQ
this can be expressed as 
 \BEQ \hspace{-3mm}
Q(r) \iss \Qc F_q(\frac{r}{R_i}) \iss m_PR_i \,\qc  \, F_q(x) , \hspace{2mm} 
 F_q(x) \iss \! \int_0^x\d y\, y^2f_q(y) ,  \hspace{2mm}  \qc  =\frac{\Qc }{m_P R_i} ,
\EEQ
with core charge $Q_c$ and with $f_q=0$ and $F_q=1$ for $x\ge1$, that is, in the mantle and the exterior.
Despite the general relativistic metric, this generates the stress energy tensor $T^\mu_{E\,\nu}=\rhoA C^\mu_\ednu$ as in special relativity,
and occurring in (\ref{TMnrhomA}), with
\BEQ\label{rhoAfq}
\rhoA (r)=\frac{E^2(r)}{2\mu_0}=\frac{Q^2(r)}{2\mu_0 r^4}=\frac{m_P^2}{8\pi  R_i^2}  \frac{\qc ^2F_q^2(x)}{x^4} .
\EEQ

Solutions for the problem will be presented in sections \ref{exactsols} and \refl{sec-col-mat}.
We first provide a physical mechanism for the local cosmological constant $\rho_\lambda$. 

\renewcommand{\thesection}{\arabic{section}}
\section{Physical nature of the zero point energy}
\setcounter{equation}{0}
\renewcommand{\theequation}{6.\arabic{equation}}
\label{zeZPE}

Before moving on to explain the term $\rhom $  in (\ref{rhoL=}), we recall some generalities.
The Casimir effect \cite{casimir1948attraction}, a geometric effect \cite{ballan2006geometry}, describes the attraction between 
two parallel conducting plates, as is 
observed  \cite{lamoreaux1997demonstration,harris2000precision,chan2001quantum,sushkov2011observation}. 
Thus by bringing them from infinity to a certain distance, energy is extracted 
from the vacuum, to be restored in the reverse action. But a conducting spherical shell has  a positive
zero point energy (ZPE) and a tendency to expand \cite{boyer1968quantum}.
The cosmological constant is generally expected to derive somehow from the ZPE of quantum fields, even though being 
much smaller than estimates thereof, and even though long believed to be exactly zero.
Would it have turned out negative, it would also be connected to ZPE, because after discarding the divergent term,
the finite part is unrestricted.

These examples show that the ZPE is set by the matter, acting as a {\it zero point battery} or a {\it zero point storage}.
For BHs  this implies that the ZPE can store part of the energy, depending on the mass distribution, hence on the metric.

\subsection{Higgs condensate and zero point energy}

From the outset, our program is to consider a suitable form for the functions $N,S$ and, next, to explain the energy momentum
tensor imposed by the Einstein equations. The $\rhoA $ term in (\ref{TMnrhomA}) was related to static electric
charges, so, with $\sigma_\vth\approx0$,  the remaining task is to explain $\rhom $. 
 Following  \cite{nieuwenhuizen2021interior} we assume a slowly varying Higgs condensate $\vph(r)$.
 It has negligible kinetic energy and the potential energy 
\BEQ \label{Vcl=}
&& \hspace{-9mm}
V(\vph)=\frac{\lambda}{4}\vph^4+\frac{m^2}{2}\vph^2+\Om=\rho_\vph+\rho_\zp  , \hspace{6mm} m^2<0,\nn  \\&& \hspace{-9mm}
\rho_\vph=\frac{\lambda}{4}(\vph^2-v^2)^2,\quad \rho_\zp=\Om-\frac{\lambda}{4} v^4 , \quad  
v^2=-\frac{m^2}{\lambda} . 
\EEQ
With these ingredients, $\rhom $ decomposes in general as 
\BEQ\label{rhomdec}
\rhom = \rho_\vph+\rho_\zp.
\EEQ
At the classical level, $\vph=v$ minimizes the potential, leaving $\rho_\zp=V(v)$ as a tuneable parameter
for equating $\rhom $ of eq. (\ref{rhoL=}) and (\ref{rhomdec}). When quantum corrections are taken into account,
an effective potential is constructed \cite{coleman1973radiative,peskin2018introduction}, 
that we represent by the form (\ref{Vcl=}), with all parameters renormalized.
The minimum is taken at the new value of $v$.
While $\Om$ now has a renormalized part \cite{ford1993effective}, the Callan-Symanzik equation allows to add any constant to it.
Since the thusly obtained $\rho_\zp$ may vary at the macroscopic scale, it remains possible to 
explain $\rhom $ by this zero point energy density.
In the approximation $\rho_\vph=\tau_\vth=0$ it  is depicted in figure 1 and and for $\rho_\vph=0$ in figure 2.

\subsection{Higgs condensate at finite $T$}

In the standard model, $\vph$ determines the particle masses, which, in their turn,  source it.
In leading order in the couplings it obeys
\BEQ\label{vph-eq}
\lambda\vph^3+m^2\vph+\frac{1}{\vph}\tau=0,\qquad \tau=\tau_B+\tau_F,
\EEQ
 the equivalent of the Gross-Pitaevskii equation for cold atoms \cite{pitaevskii2016bose}.
 $\tau_B$ stems from the bosons $b=H$, $Z$, $W$,   and $\tau_F=\sum_\itf \tau_\itf$ from all fermions,
 $\itf \iss u,d,e,\nu_e$; $c,s,\mu,$$\nu_\mu$ and $t,b;\tau,\nu_\tau$.
In a thermal state, the boson term reads at lowest order in perturbation theory (loop expansion)
 \BEQ \label{tauB=}
 \hspace{-3mm}
 \tau_B \iss \!\int \! \frac{\d^3k}{(2\pi)^3} \! \left(
\frac{f_H\plus \half}{k_0/ 3\lambda\vph^2}
 \pplus 3\frac{f_Z\plus \half}{k_0/M_Z^\vph{}^2}
\pplus 6\frac{f_W\plus\half}{k_0 /M_W^\vph{}^2}\right )\! .
 \EEQ
 Here $f_b$ the Bose-Einstein distribution $1/(e^{\beta (k_0-\mu_b)} -1)$ and $\half$
  the zero-point term; $k_0^2=k^2+M_b^\vph{}^2$ with $M_H^\vph{}=\sqrt{\lambda(3\vph^2-v^2)}$ 
  and $M_b^\vph=M_b\vph/v$ for $b=Z,W$.
For fermion $\itf$ one has, with  $m_\itf^\vph =m_\itf \vph/v$  and $k_0^2=k^2+m_\itf^\vph{}^2$, 
 \BEQ \label{tauf=}
 \hspace{-3mm}
 \tau_\itf \iss
 \! \! \int \! \! \frac{\d^3k}{(2\pi)^3}\frac{g_\itf m_\itf^\vph{}^2 }{2\hspace{0.3mm} k_0} \! \! \left(
\frac{1} {e^{\beta( k_0-\mu_\itf)}\pplus 1}\pplus \frac{1} {e^{\beta (k_0-\bar\mu_\itf)}\pplus 1}\mmin 1 \right) \! ,
\EEQ
with $g_\itf=12$ for quarks, $4$ for $e$, $\mu$, $\tau$, and $2$ for neutrinos, and
chemical potential  $\mu_\itf$ ($\bar\mu_\itf$) of the (anti)fermion $\itf$.
The zero point terms lead to the renormalization of the effective potential $V(\vph)$ and should further be skipped.
In a thermal state, $T$ and the various $\mu$ are functions of $r$, while Higgs particles can form a condensate ($\vph(r)\neq v$).

\renewcommand{\thesection}{\arabic{section}}
\section{Exact solutions for the metric}
\setcounter{equation}{0}
\renewcommand{\theequation}{7.\arabic{equation}}

\label{exactsols}

We first consider the zero-temperature limit $T(r)=0$, where no thermal particles are generated and
we postpone inclusion of the rest  masses of the collapsed up and down quarks and electrons to the next section.

\subsection{General charge distribution in the core}

Approximating $\sigma_\vth = \tau_\vth=0$ allows the simplification $\Sn (r)=1$.
Equating (\ref{rhoA=}) and (\ref{rhoAfq}) yields
\BEQ\label{rhoAfq2}
\rhoA (r)=\frac{m_P^2}{8\pi  R_i^2}  \frac{\qc ^2F_q^2(x)}{x^4}  ,
\qquad x=\frac{r}{R_i} .
\EEQ
 Writing from now on $\Se  (r)$ as $\Se  (x)$, this  combines with (\refl{rhoA=}) into
\BEQ  \label{S1xeqn}
2\Se  (x)-x^2\Se  ''(x)=4q_i^2 \frac{F_q^2(x)}{x^2}. 
\EEQ
The solution with $\Se  (r=R_i)=\Se  (x=1)=1$ reads
\BEQ \label{S1exactgeneral} \hspace{-3mm}
\Se  (x) =x^2\big\{1+\frac{4}{3}\qc ^2\left[{}J_q(1)-J_q(x)\right] \big\},\hspace{1mm}
J_q(x)= \int_0^x \d y( \frac{1}{y^3}-\frac{1}{x^3})\frac{F_q^2(y)} {y^2}  .
\EEQ
A finite charge density at $r=0$ leads to $F_q(y)\sim y^3$ for $y\to0$, so the integrals are well behaved.
It follows that
\BEQ\label{si-Iq}
s_i^-=R_i\Se  '(R_i^-)=2-4\qc  ^2I_q,\quad I_q=\int_0^{1}\d y\frac{F_q^2(y)}{y^2} .
\EEQ
So $s_i^-\le 2$, while $s_i^- > 0$ is required at the first crossing of $\Se  =1$ starting from $S(0)=0$.
For $\Sn (r)=1$ eq. (\ref{rhoL=}) takes the form
\BEQ
\rhom  =\frac{2\Se  \pplus 4x \Se  '\pplus x ^2\Se  ''}{32\pi G x^2}
\EEQ
At $R_i$, corresponding to $x=1$, one has
\BEQ\label{rhoLR<}
\rhom (R_i^-)=\frac{3-\qc ^2 -4\qc ^2I_q}{8\pi GR_i^2} .
\EEQ
Continuity with the vacuum in the mantle fixes
\BEQ\label{qisqsi}
\qc ^2 =\frac{3}{1+4I_q} =1+s_i^-  .\qquad 
\EEQ
The condition $s_i^->0$ at the first crossing of $\Se  =1$, starting from $\Se  =0$ at $x=0$,
requires that $I_q<\half$, which constrains allowable charge distributions.
Continuity $s_i^-=s_i^+=R_i\Se  '(R_i^+)= 2p_c/(1-p_c)$  with $p_c=\sqrt{1-\cQ_c^2/M_c^2}$   sets
\BEQ
q_c\equiv \frac{\cQc   }{\Mc }=\frac{2\qc }{1+\qc ^2}=\frac{\sqrt{3(1+4I_q)}}{2(1+I_q)}
,\quad
R_i=\frac{2G\Mc }{1+\qc ^2} ,
\quad
 R_e=\frac{2\qi^2G\Mc }{1+\qc ^2} .
\EEQ
With $s_i^-$ between $0$ and $2$ due to (\ref{si-Iq}), $q_c$ ranges from  $\half \sqrt{3}$ to $1$, that is to say, from quite charged to
maximally charged.  

\subsection{Constant and linearly decaying charge density}

Realistic cases exist.
A uniform charge density has $f_q(x)=3$, $F_q(x)=x^3$, $I_q={1}/{5}$, $\qc =\sqrt{5/3}$, $s_i=2/3$ 
and $q_c=\cQc   /\Mc =\sqrt{15}/4$. 

More realistic is the charge density
\BEQ \label{goodQr}
\rho_q=\frac{15\Qc }{8\pi R_i^3}(1-x^2),
\EEQ
which vanishes at $r=R_i$. 
(We skip the $1-x$ case, since the absence of odd powers of $x$ simplifies the ensuing expressions.)
This leads to the properties
\BEQ \label{goodQx}
&& \hspace{-4mm}
\Se  (x)=\frac{x^2}{38} \left [105(1- x^2) + 45 x^4 - 7 x^6\right], \quad \\
&& \hspace{-5mm}
F_q(x)=\half (5-3x^2)x^3,\quad I_q=\frac{3}{7},\quad \qc  =\sqrt{\frac{21}{19}}, 
 \quad s_i=\frac{2}{19} , 
\nn
\\ && \hspace{-4mm}
\frac{\cQc   }{\Mc }=\frac{\sqrt{399}}{20},
\quad \frac{R_i}{G\Mc }=\frac{19}{20},\quad \frac{R_e}{G\Mc }=\frac{21}{20}. 
\nn
\EEQ
The charge to  mass ratio is closer to 1 than in the uniform case, since the charges are more centered,
which enhances $I_q$ and lowers $\Qc $.
The energy density $\rhom $ decreases monotonically and vanishes quadratically at $R_i$, 
\BEQ \label{cTexact}
\rhom  = \frac{m_P^2}{32\pi R_i^2 }\frac{315}{19}(2-x^2)(1-x^2)^2 . 
\EEQ
This function is plotted as the upper curve in figure 1, 
with the  related $\rhoA $, defined by (\ref{rhoAfq}) with $F_q=1$ for $x\ge 1$.

 \begin{figure}\label{figrhozpRiRe} 
\centerline{ \includegraphics[width=8cm]{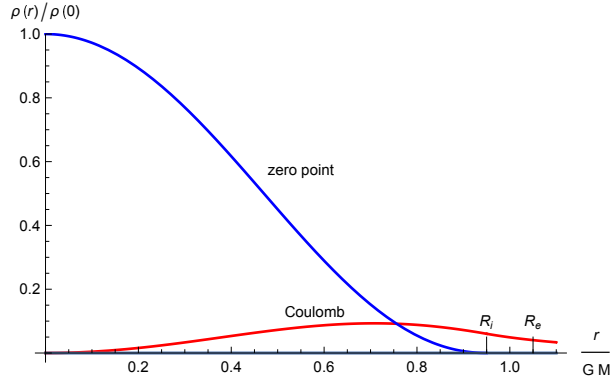}}
\caption{ In a charged black hole,  matter is located inside the core bounded by the inner horizon $R_i$.
At zero temperature of the matter, the energy is stored as zero point energy of the quantum fields (upper curve, for the profile  (\ref{cTexact})).
Like the outer space, the mantle between $R_i$ and  the event horizon $R_e$
is a vacuum described by the Reissner Nordstr\"om metric.
The electrostatic energy density (lower curve, eq. (\refl{rhoAfq}) for the charge distribution (\ref{goodQr}))
has a Coulomb tail $Q_c^2 \! /8\pi r^4$ outside the core.
}
\end{figure}

The total $\Lambda$-energy is
\BEQ
\int_0^{R_i}\d r \, 4\pi r^2\rhom =\left.\frac{r(2\Se  +r\Se  ')}{8G} \right\vert_0^{R_i}
 =\frac{5R_i}{19G}
=\frac{\Mc }{4},\quad
\EEQ
while the Coulomb energy adds
\BEQ
\int_0^\infty \d r\,4\pi r^2\rhoA =\left. \frac{r(2\Se  -r\Se  ')}{8G}\right\vert_0^\infty =\frac{3}{4}\Mc  .
\EEQ
So the total mass (energy) is $\Mc $. While the mantle contribution is involved despite its reversed role of $r$ and $t$,
this mass is confirmed \cite{nieuwenhuizen2021interior} via a field theoretic approach related to 
the Landau Lifshitz pseudo tensor \cite{landau1975classical,babak1999energy,nieuwenhuizen2007einstein}.

\subsection{Quadratically decaying charge densities}
\label{quad-decay}

When taking into account the thermal matter in section \ref{sec-col-mat}, we shall need that
 $\sigma_\vth$ vanishes at least as $(r-R_i)^2$. It is then natural to assume that also the charge density does so.
But not every case works. 
For instance, $f_q(x)=105(1- x^2)^2/8$ leads to $I_q=271/429>\half$ and the unphysical $s_i^-=-226/1513$.

Hence, we consider still keeping $\Sn =1$ and $\sigma_\vth=0$, 
\BEQ \hspace{-3mm} \label{secondfq}\label{fq-casea}
\rho_q = \frac{Q_cf_q(x)}{4\pi R_i^3} , \qquad f_q=\frac{231}{32}(1-x^4) ^2,
\EEQ
for which the normalized included charge and its $I_q$ read
\BEQ\label{Fqquad}\label{Fq-casea}
F_q=x^3\frac{ 77 - 66 x^4 + 21 x^8}{32},\qquad I_q=\frac{1613}{3315}=0.486576.
\EEQ
Its $s_i^-=178/9767=0.01822$ implies a nearly maximal charge,  $1-\cQ_c/\Mc =4.1\,10^{-5}$.
Eq. (\ref{S1exactgeneral}) leads to the exact form
\BEQ  \label{S1ex3rd}\label{S1-casea}
\Se  =\frac{109395 x^2}{39068}\left (
1-\frac{539 x^2}{640}+\frac{77 x^6}{288} -\frac{69 x^{10}}{832}
+\frac{9 x^{14}}{544}-\frac{7 x^{18}}{4224} \right) .
 \EEQ
Equation (\ref{rhoL=}) for this case with $\Sn =1$ reads
 \BEQ   \label{rhomex3rd}
  \hspace{-3mm} \rhom= \frac{109395}{2500352} \frac{768 - 1617 x^2 + 1540 x^6 - 966 x^{10} + 324 x^{14} - 49 x^{18}}{32\pi G R_i^2} ,
  \EEQ
which contains a factor $(1-x^2)^3$. It is depicted  in fig. 2, with $\Se  $ of eq. (\ref{S1ex3rd}).

So far we considered a finite charge density at the origin.
A case in which this term vanishes there quadratically and also at the inner horizon, is
\BEQ\label{fq-caseb}
f_q(x)=\frac{315}{8}x^2(1-x^2)^2, \quad  F_q(x)=\frac{x^5}{8}  (63 - 90 x^2 +35 x^4).
\EEQ
 It involves $s_i=\Se  '(1)=382/2049=0.186432$ and more fully
\BEQ &&
\hspace{-9mm} 
\label{S1-caseb}
\Se  \iss
\frac{119119}{65568 }x^2 \left(1 - 3 x^6 + \frac{405}{77} x^8- \frac{2502}{637}x^{10} + \frac{10}{7} x^{12} - 
\frac{ 25}{119} x^{14} \right ).
\EEQ
The local cosmological constant  decays monotonically,
\BEQ
\rho_\lambda=\frac{17017(1 - x^2)^3 }{10928\times 32\pi GR_i^2} (14 + 42 x^2 + 84 x^4 - 175 x^6 + 75 x^8) .
\EEQ

Another case has a charge density $\sim r^4$ near the origin. It is described by
\BEQ\label{fq-casec}
f_q(x)=\frac{693}{8}x^4(1-x^2)^2 , \qquad  F_q(x)=\frac{x^7}{8}  (99 - 154 x^2 + 63 x^4),
\EEQ
and 
\BEQ \hspace{-9mm} \label{S1-casec}
\Se   &\iss& \frac{1339481 x^2}{942912} \! \left(
1 \mmin \frac{1782 }{377}x^{10} \pplus \frac{308 }{29}x^{12} \mmin \frac{4700 }{493}x^{14} \pplus \frac{2205 }{551}x^{16} \mmin \frac{210}{319} x^{18} \right) \! .
\EEQ
It involves $\Se  '(1)=6262/14733=0.425032$.
The local cosmological constant 
\BEQ
\rho_\lambda=\frac{46189}{78576}  \frac{29 - 2079 x^{10} + 6160 x^{12} - 7050 x^{14} + 3675 x^{16} - 735 x^{18}}{32\pi G R_i^2} ,
\EEQ
again decays cubically, since also this form contains a factor $(1-x^2)^3$.

The charge-to-mass ratio $q_c=\cQ_c/M_c=2q_i/(1+q_i^2)$ for these three cases takes the values
$1-q_c=4.07716 \,10^{-5}$, $3.64194\,10^{-3}$ and $1.54794\,10^{-2}$, respectively, so $q_c$ is smaller 
when the charges are further away from the origin.

\subsection{All charges located near the inner horizon}
\label{toRi-}

 From eq. (\ref {qisqsi})  the maximum of $\qi$ follows as  $ \sqrt{3}$,  and occurs for $I_q\to  0$. Let us consider this limit for the case
\BEQ
f_q(x)=\frac{ 3 \eps +x^2}{\eps}e^{-(1-x^2)/2\eps} ,\qquad F_q(x)=x^3 e^{-(1-x^2)/2\eps} ,
\EEQ
where, for simplicity, we allow a  finite $f_q(1)$.  It leads to
\BEQ
I_q=\int_0^1 \d x\, \frac{F_q^2(x)}{x^2} =\frac{\eps}{2} + \frac{3\eps^2}{4} [K(1)-1], \quad K(x)= \int_0^x\d x\, e^{-(1-y^2)/\eps} .
\EEQ
The solution for $\Se  $ reads
\BEQ
\Se  =x^2+\eps^2\qi^2 \left[ \, x^2  -e^{-(1-x^2)/\eps} +\frac{1}{x}K(x) -x^2K(1)\,\right] ,
\EEQ
satisfying $\Se  =1$ at $R_i$  ($x=1$). With
   \BEQ    \qi^2=\frac{3}{1+2\eps+3\eps^2[K(1)-1] } 
   \EEQ
the local cosmological constant vanishes at $x=1$,
\BEQ 
\rho_\lambda=\frac{3m_P^2}{8\pi R_i^2} \frac{ 1+2\eps-  (x^2+2 \eps) e^{-(1-x^2)/\eps} }{1+2\eps+3\eps^2[K(1)-1]} .
   \EEQ

   In the limit $\eps\to 0$ the charges get ``pushed'' towards the inner horizon. This causes that $I_q\to0$, so that $\qi\to\sqrt{3}$ and $s_i^-\to2$. 
At a given $Q_c$ and still neglecting the zero point masses, the maximal BH mass that can be allowed emerges 
as $M_c=2\cQ_c/\sqrt{3}$, or the minimal allowed BH charge-to mass ratio as $q_c=\cQ_c/M_c=\frac{1}{2}\sqrt{3}$.

In the limit $\eps\to0$, at fixed $x<1$ and $\Sn =1$, we have a solution with an infinitesimal charge layer 
and a discontinuous $\rho_\lambda$. It has the simple form
\BEQ
\Se  =x^2,\quad \rho_\lambda =\frac{ 3m_P^2}{8\pi R_i^2}\theta(R_i^--r),
\quad \rho_q=\frac{Q_c}{4\pi R_i^2}\delta(r-R_i^-) .
\EEQ
where $\theta$ is the Heaviside step function. This implies $s_i^-=2$ and
\BEQ
\hspace{-6mm}
Q(r)=Q_c\theta(r-R_i^-),\quad \cQ_c=m_PQ_c=\frac{\sqrt{3}}{2}M_c,\quad \rhoA =\frac{Q_c^2}{8\pi r^4}\theta(r-R_i^-).
\EEQ
Fitting $\Se  '$ at $R_i$, where $\Se  =1$,  to a Schwarzschild metric in the mantle
is impossible since it has $\Se  '<0$;  the RN metric is needed, with 
\BEQ
S=2\frac{GM_c}{r}-\frac{GQ_c^2}{r^2} ,\qquad rS'=-2\frac{GM_c}{r}+2\frac{GQ_c^2}{r^2} 
\EEQ
The inner and event horizons lie respectively at
\BEQ\label{Ri2Ri}
\hspace{-5mm}
\frac{R_i}{G}=M_c-\sqrt{M_c^2-\cQ_c^2}=\half M_c,
\hspace{1mm}
\frac{R_e}{G}=M_c+\sqrt{M_c^2-\cQ_c^2}=\frac{3}{2}M_c.
\EEQ
The energy densities $\rho_\lambda$ and $\rhoA $ lead to the mass contributions
\BEQ  \hspace{-6mm}
\int_0^{R_i}\d r\,4\pi r^2 \rho_\lambda=\frac{R_i}{2G}=\frac{1}{4}M_c,\quad 
\int_{R_i}^\infty\d r\,4\pi r^2 \rhoA =\frac{Q_c^2}{2R_i}=\frac{3R_i}{2G}=\frac{3}{4}M_c.
\EEQ
Their sum equals $M_c$, as expected.

\subsection{Partially negative charge density}

There are solutions in which the net charge density is negative in the outskirts, for example
\BEQ
f_q(x)=\frac{14}{3} x^2 (25 - 26 x) (1 - x),\qquad F_q(x)=\frac{1}{3} x^5 (70 - 119 x + 52 x^2).
\EEQ
It involves $f_q<0$ between 25/26 and 1. This case leads to
\BEQ \hspace{-3mm}
S(x)=x^2\frac{ 56749 - 1078000 x^6 + 2827440 x^7 - 2894535 x^8 + 1361360 x^9 -  247104 x^{10} }{ 25910} ,
\EEQ
which involves $s_i^-=82/2591 = 0.031648 $. Interestingly, the local cosmological constant
\BEQ
\rho_\lambda= 231 \frac{737 - 105000 x^6 + 336600 x^7 - 413505 x^8 + 229840 x^9 - 48672 x^{10} } {207280\, \pi G R_i^2} ,
\EEQ
is negative between $x_0\equiv 0.942308$ and $1$. The related energy is moved inwards. The total energy in the local cosmological constant
reads
\BEQ
\int_0^{R_i}\d r\,4\pi r^2 \rho_\lambda=\frac{658\, R_i}{2591\, G}.
\EEQ
It is located in the core region $x\le x_0$, since an additional
fraction of 0,0001414 of zero point energy is moved from the core outskirts $x_0<x<1$ to the core interior $x<x_0$.

This example show that the local cosmological constant can be negative in some region.

\renewcommand{\thesection}{\arabic{section}}
\section{Effect of the collapsed matter at $T=0$}
\setcounter{equation}{0}
\renewcommand{\theequation}{8.\arabic{equation}}

\label{sec-col-mat}

In our $T=0$ limit, bosons are absent but the collapsed up and down quarks and electrons remain, 
carrying 1\% in rest mass compared to the liberated nucleon binding energy.
They reside in their quantum ground state with Fermi energies $\eps_{u,d,e}$. 
In the local Minkowski frame, the latter have the properties
\BEQ
n_e =  \!\int_{\!<k_e} \!\!\frac{\d^3k}{4\pi^3}=\frac{k_{e}^{3}}{3\pi^2}, \qquad
(\rho_e,p_e) = \! \int_{\!<k_e} \!\!\frac{\d^3k}{ 4\pi^3}(k_0,\frac{k^2}{3k_0}) ,
\EEQ
with $k_0=\sqrt{k^2+m_e^2}$. The total number of electrons is $N_e=4\pi\int_0^{R_i}\d r\,r^2n_e\approx \Mc /m_N$.
It follows that 
 \BEQ \hspace{-4mm}
\sigma_e&=&\rho_e+ p_e=n_e\eps_e,\quad \eps_e= \sqrt{k_e^2+m_e^2} ,
 \\  \hspace{-4mm} \tau_e&=&\rho_e-3p_e=\frac{m_e^4}{2\pi^2}\left (\frac{k_e\, \eps_e}{m_e^2}- {\rm arcsinh}\, \frac{k_e}{m_e}\right) .
\EEQ
Similar forms hold for the $u$ and $d$ quarks, with $N_u\approx 2N_e$ and $N_d\approx N_e$
set by the metal content of the precollapse matter.
They add up to $\sigma_\vth=\sigma_u+\sigma_d+\sigma_e$ and $\tau_\vth=\tau_u+\tau_d+\tau_e$,
and are relativistic for $\Mc \lesssim 10^4M_\odot$.
At $T=0$ the contribution $\rho_\vph$ is negligible since $\tau=\tau_\vth\ll v^4$.

 The function  $\Sn (r)$ is no longer trivial. 
The condition $\Sn '(R_i)=0$ is needed in (4) to achieve  $\rhom (R_i)=0$, owing to (\ref{rhoAfq}) and (\ref{qisqsi}). 
It demands via (\ref{sigth=}) that $\sigma_\vth$ vanishes  at least as $(r-R_i)^2$.  Hence we reconsider the charge distribution 
(\ref{secondfq})
of subsection \ref{quad-decay} and assume, for simplicity, a constant charge-to-mass ratio\footnote{The local charge-to-mass ratio can 
be seen as an ``equation of state''.
Since the net charge density $\rho_q$ embodies a only tiny fraction of the electron charge density,  
viz. $\rho_q/en_e=Q_cm_N/eM\le m_N/em_P=9\,10^{-16}$,  non-uniform ratios are realistic.}
so that the densities and total numbers of up and down quarks and electrons are given by
\BEQ \hspace{-3mm} \label{secondfq2}
\frac{n_u}{N_u}=\frac{n_d}{N_d}=\frac{n_e}{N_e}=\frac{\rho_q}{\Qc }=\frac{f_q(x)}{4\pi R_i^3}=
\frac{231}{32}\frac{(1-x^4) ^2}{4\pi R_i^3} . \hspace{1mm}
\EEQ
To solve (\ref{rhoA=}), (\ref{sigth=}) with $\sigma_\vth$ included for the case (\ref{secondfq2}), 
and determine  $\rho_\lambda$ from eq. (\ref{rhoL=}), 
one moves from the function $\Sn $ to $U=-\Sn ^2\bar \Se  $ to eliminate $\Se  ''$, 
and eliminates the bilinear $U'{}^2$ and $U'\Se  '$ terms in a certain linear combination of (\ref{rhoA=}) and (\ref{sigth=}).
Series expansion in powers of  $x^2$ leaves $\qc $ and $u_2=U''(0)/U(0)$ as free parameters.
Near $R_i$ one can set $x=1-y^3$ and expand in $y$. Since $\Se  '(R_i^-)=(\qc ^2-1)/R_i$, there appears no new free parameter.
Integrating inwards from both ends allows to fix $\qc $ and $u_2$ by matching $U'/U$ and $\bar \Se  $;
 lastly, the ratio $\Sn =\Sn (0)/\Sn (R_i)$ is read off by matching $U$. The results are as follows:
For $\Mc \gg 10^4M_\odot$  it holds that $s_i^-=0.01815$ and $\Sn =0.999704$ instead of 1; 
for $\Mc =2M_\odot$ this becomes $s_i^-$ = $0.0051401$, and $\Sn =0.9554$. 
So inclusion of the  ground state energies 
has a 5\% effect at worst and makes the enclosed charge nearer to maximal.   

The influence of the collapsed matter is depicted  in fig. 2.

 \begin{figure}\label{figrhozpRiRe} 
\centerline{ \includegraphics[width=8cm]{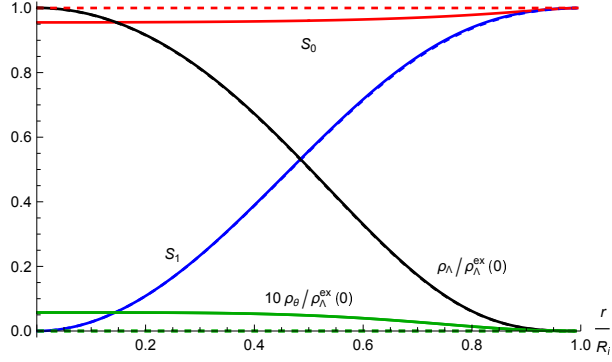}}
\caption{
Dashed lines are the exact (``ex'') radial profiles of the metric functions $\Sn (r)=1$ and $\Se  (r)$ from eq. (\ref{S1ex3rd});
 the zero point energy density  $\rho_\lambda^\ex(r)$ of (\ref{rhomex3rd}) normalized to $\rho_\lambda^{\rm ex}(0)$, 
 and $\rho_\vth(r)=0$,  for the charge distribution (\ref{secondfq}).  
The solid lines present these functions when accounted for the rest mass of the collapsed up and down quarks and electrons for a  core mass 
of $2M_\odot$, for  $\Sn (R_i)=1$.
The lower curve is the fermionic ground state energy density $\rho_\vth$  ($\times 10/\rho_\lambda^{\rm ex}(0)$); 
unlike for $\Sn $, its effect on $\Se  $ and $\rho_\lambda$ is nearly invisible.
}
\end{figure}

\renewcommand{\thesection}{\arabic{section}}
\section{Perturbation of the metric and electric field}
\setcounter{equation}{0}
\renewcommand{\theequation}{\thesection.\arabic{equation}}

\label{sec-stab}

Given the general expectation that all BH solutions go in the course of time to one of the standard metrics (Schwarzschild, Reissner-Nordstr\"om, Kerr, Kerr-Newman),
an intriguing question is whether our solutions expose some kind of stability. Here we lay the  groundwork for investigating this.

 \subsection{First order perturbation equations}

Consider spherically symmetric perturbations  of the metric and the electromagnetic potential,
\BEQ \hspace{-5mm}\label{deltagmn}
\delta g_{\mu\mu}=2g_{\mu\mu}h_\mu e^{-i\om t}, \quad (\mu=0,1,2,3;\,h_3=h_2),
\quad \delta A_0=A_0^\heen e^{-i\om t} ,
\EEQ
where $h_{0,1,2}$ and $A_0^\heen$ are small, bounded functions of $x=r/R_i$. It holds that
\BEQ\label{delS0=}
\frac{\delta \Se  }{\Se  }=-2h_1e^{-i\om t} , \quad
\frac{\delta \Sn }{\Sn }=(h_0+h_1)e^{-i\om t} .
\EEQ
For continuity at the core boundary  it is desired that $h_{0,1,2}$ vanish at $r=R_i$.

The aim is to search for eigenfrequencies $\om_i$; 
given that $\om^2$ will be real, any  $\om_i^2<0$ would connect to instability.
The Maxwell equations lead to
\BEQ\label{rhoA1em}
\hspace{-6mm}
\delta\rhoA =\rhoA ^\heen e^{-i\om t},\quad \rhoA ^\heen 
=\frac{A_0^\heen{}'-(h_0+h_1)A_0'}{4 \pi  \Sn ^2}A_0'
=\frac{E^\heen-(h_0+h_1)E}{4 \pi }E ,
\EEQ
where $E^\heen=-A_0^\heen{}'$ is the spatial perturbation of the electric field.
The Einstein tensor attains off-diagonal elements $G^1_{\ed0}=-\Sn ^2\bar \Se  ^2G^0_{\ed1}\sim\om$. 
They cannot be accommodated by the Anstatz (\ref {TMnrhomA}) nor by allowing elements $\delta g_{01}=\delta g_{10}=h_{01}(r)\exp(-i\om t)$, 
but they drop out by imposing
\BEQ\label{h1r=}
h_1=r h_2'+h_2-h_2 \left( \frac{r\Se  '}{2 \bar \Se  } +\frac{r\Sn '}{\Sn } \right) .
\EEQ
The remaining Einstein equations correspond to the first order perturbations in the coefficients of  (\ref {TMnrhomA}), that take the form
\BEQ
\delta\rho_\lambda=\rho_\lambda^\heen(r) e^{-i\om t},  \quad
 \delta\rhoA =\rhoA ^\heen(r) e^{-i\om t} ,\quad
\delta\sigma_\vth=\sigma_\vth^\heen(r)  e^{-i\om t} ,
\EEQ
In terms of the dimensionless radius $x=r/R_i$, the spatial parts take the forms
\BEQ
&& \hspace{-3mm}
8\pi GR_i^2\rho_\lambda^\heen \! = \!
 \left( \frac{\ombar ^2}{\bar \Se  }\! -\! \Se  ''-\frac{4 \Se  '}{x}-\frac{2 \bar \Se  }{x^2}\right)\frac{h_1}{2} 
\! +\!   \frac{\ombar ^2x^2-1}{\bar \Se  x^2}h_2 
\! + \!  \frac{\Se  '}{4}  \left(3 h_0'-h_1' +4 h_2'\right)
 \nn\\&&
+ \, \frac{   \left(2 h_0'-h_1'+2 h_2'\right) x \bar \Se  -4 h_1 \bar \Se  -3 x h_1 \Se  ' }{2 x } \frac{\Sn '}{\Sn }-\frac{\bar \Se   h_1 \Sn ''}{ \Sn } ,
\EEQ
where $\ombar (r)=\om R_i /\Sn (r)$, and 
\BEQ  \label{rhoA1} \hspace{-6mm}
8\pi GR_i^2\rhoA ^\heen &=&
\frac{\ombar ^2 x^2-2 \bar \Se  }{2 x^2 \bar \Se  }h_2 -\frac{\ombar ^2 x^2-x^2 \bar \Se   \Se  ''
+2 \bar \Se  ^2}{2 x^2 \bar \Se  }h_1
\nn\\ \hspace{-6mm}
&+&\, \frac{ \bar \Se  }{2 } \left (\frac{h_0'+h_1'}{x}-h_0''-h_2'' \right)
+\frac{\Se  ' }{4} \left(h_1'-3 h_0'\right)
 \nn\\  \hspace{-6mm}
&+& \, \left[\frac{3}{2}  h_1 \Se  '-\frac{ 2 h_1-x \left( h_1'+h_2'-2 h_0' \right)}  {2 x} \bar \Se   \right]\frac{\Sn '}{\Sn }  
+\frac{\bar \Se   h_1 \Sn ''}{ \Sn } ,
\EEQ 
and
\BEQ 
\label{sig1} \hspace{-6mm}
8\pi GR_i^2\sigma_\vth^\heen= \frac{3 \bar \Se   \left(2 h_1-xh_2'\right)}{2 x} \frac{\Sn '}{\Sn } 
-\frac{3 \ombar ^2 h_2}{2 \bar \Se  }  -\frac{3 \bar \Se   \left(h_0'+h_1' -2 h_2'- x h_2'' \right)}{2 x} .
\EEQ

\subsection{Perturbation of the charge distribution}
\label{pertcharges}

The exact solution of section \ref{exactsols} was set by the charge distribution, so one expects this to carry over to the perturbations.
This is indeed shown now.

The incorporation of the $f=u,d,e$ rest masses in section \ref{sec-col-mat} is an example of a perturbation of the leading order exact solution
of section \ref{quad-decay}. While that analysis pertains to frequency $\om=0$, we proceed in the same spirit for general $\om$.
As before, we take $T=0$, $\sig_\vth^ \hnul=0$ and $\Sn =1$. $\Se  ^\hnul (x)\equiv \Se  (x)$ satisfies
\BEQ\label{S1xeqn}
2\Se  -x^2\Se  ''=4q_i^2 \frac{F_q^2}{x^2}.
\EEQ
 For simplicity, we restrict ourselves to large BH mass where $k_\itf  \ll m_\itf $  and $\sig_\vth^\heen=\sum_\itf m_\itf n_\itf^\heen=\mu M$.
With $\half N_u=N_{d,e}=M/m_N$ this yields 
\BEQ \hspace{-6mm}
\sig_\vth^\heen \iss\frac{M}{Q_c} \mu\rho_q^\heen \iss \bar\mu \rho_q^\heen
, \hspace{2mm} \bar\mu=(1 \pplus q_i^2) \mu, \hspace{2mm} \mu=\frac{2m_u \pplus m_d \pplus m_e}{m_N}=9.28\,10^{-3} .
\EEQ
 For some profile $\varrhoq(x)$ with max($\varrhoq)=1$ we assume a small perturbation of the enclosed charge
\BEQ\label{varrho}
Q_c^\heen(x)=\eps Q_c\varrhoq(x)F_q(x),\quad   
\EEQ
where $\eps\ll 1$. The underlying perturbation of the charge density is
 \BEQ \label{tildefq}
\rho_q^\heen(x)=\frac{\eps Q_c}{4\pi R_i^3} \fqrho (x),  
\qquad  \fqrho (x) =   \varrhoq(x)f_q(x)+ \varrho_q'(x) \frac{F_q(x)}{x^2} .
\EEQ
The total charge $Q_c+\eps Q_c^\heen(1)e^{-i\om t}$ should be conserved for fluctuations of the matter 
inside the core, which imposes $\varrhoq(1)=0$ in (\ref{varrho}). 
Since the $h_{0,1,†2}$ are proportional to $\eps$, we can set $\eps=1$ from now on.

With $E^\heen=Q_c^\heen/r^2=\varrhoq E$ we get from (\ref {rhoA1em})
\BEQ  \label{rhoAsource}
\hspace{-6mm}
\rhoA ^\heen = q_i^2 \frac{\varrhoq (x)-h_0(x)-h_1(x) } {4\pi GR_i^2}\frac{F_q^2(x)}{  x^4}  .
\EEQ

For continuity, $\rhoA ^\heen$ has to vanish at $R_i$, so that $h_0+h_1=\varrhoq $ at $x=1$.
With $h_0=h_1=0$ at $x=1$, as discussed after (\ref{delS0=}), this confirms that $\varrhoq(1) =0$ .

The differential equations for $\rhoA ^\heen$ and $\sig_\vth^\heen $, originating from the Einstein equations, are formally of second order.
However, series expansion around $x=0$ learns that there are only 2 integration constants.
Indeed, it appears possible to combine these equations  into a single second order differential equation. In
\BEQ \label{sig3} 
\hspace{-6mm} &&
8\pi GR_i^2\sigma_\vth^\heen= -\frac{3 \ombar ^2 h_2}{2 \bar \Se  } 
-\frac{3 \bar \Se  }{2x} \left(h_0+h_1-(xh_2)'\right)'=\bar\mu\fqrho 
\EEQ
we  eliminate $h_0$ in favor of a ``nucleus'' $\gn (x)$ with $g(1)=0$, and employ (\ref{h1r=}), 
\BEQ
h_0=\gn +\frac{x\Se  '}{2 \bar \Se  }h_2, \qquad
\gn=h_0+h_1-(xh_2)' .
\EEQ
Equation (\ref{sig3}) now allows to solve for $h_2$, which then determines $h_0$ and $h_1$,
\BEQ \label{h012g}
h_2 &=& -\frac{ \bar \Se  ^2} { \ombar ^2 x} \gn '  -\frac{ 2\bar\mu\bar \Se  } {3 \ombar ^2 } \fqrho ,\qquad
h_0=\gn -\frac{ \bar \Se  \Se  '} { 2\ombar ^2 } \gn ' - \frac{ \bar\mu x \Se  '} {3 \ombar ^2 } \fqrho , \nn\\
h_1&=& -\frac{\bar \Se   (3 \Se  ' \gn '+2 \bar \Se   \gn '' )}{2 \ombar ^2}
-\bar\mu\frac{ x \fqrho  \Se  '+2 \bar \Se   (\fqrho +x \fqrho{} ' )}{3 \ombar ^2} .
\EEQ

Equating (\refl{rhoA1}) and (\ref{rhoAsource}) for $\Sn =1$, using these relations and eliminating $\Se$ by differentiating (\ref{S1xeqn}),
yields a second order differential equation for $\gn $, 
\BEQ && \hspace{-12mm} \label{g0xeqn}
 \bar \Se  ^2 \gn '' +2 \bar \Se   \Se  ' \gn '   +\frac{1}{3x} \left(\frac{xF_q'}{F_q}-2 \right)\bar \Se  ^2 \gn ' - \ombar ^2  \gn  + G(x) =0
\EEQ
with the source $G$ involving $\varrhoq$ and $\fqrho =(\varrhoq F_q)' / x^2$ introduced in (\ref{tildefq}), 
 \BEQ \label{sourceG}
&& G=G_0 +\mu (G_1+ G_2), \quad
G_0=  \ombar ^2 \varrhoq , \quad 
G_1=   -\frac{ \ombar ^2x^5 }{12 q_i^2F_q^2}  (\frac{\Se  ' }{\bar \Se  }\fqrho+2 \fqrho' ) ,
  \nn\\&&
G_2=\frac{2}{9} [3 x  \fqrho  \Se  '+ (1+\frac{x^3 f_q}{F_q}) \fqrho \bar \Se   +3 x \fqrho ' \bar \Se  ] .
\EEQ
The left hand side of (\ref{g0xeqn}) is determined by the charge profile in the core, while the right hand side is set by the fluctuations covered in $\varrhoq$.

Since $\mu$ is small, we wish to neglect $G_{1,2}$, keeping the leading order $G_0$. While $G_2$ is not problematic,
$G_1$  remains finite  under conditions.  
The $x\to0$ behavior $f_q\sim x^q$  requires that $\varrhoq\sim x^a$, $\fqrho\sim x^{a+q}$  with $a\ge q+2$.  The $x\to1$ decay
$f_q\sim(1-x)^2$ needs to be accompanied by $\varrhoq\sim(1-x)^c$ with $c>2$.

For reasons that become clear soon, it is advantageous to express the frequency $\om$ in a parameter $p$ by
employing $s_i=R_i\Se  '(r=R_1)=\Se  '(x=1)$,
\BEQ \label{om2p}
\hspace{-6mm}
\om=\frac{\ombar }{R_i}
,\quad 
\ombar =s_i\om_p,\quad \om_p=\sqrt{p(p+1)},\quad 
p=\frac{\sqrt{1+4\ombar ^2/s_i^2} -1}{2} .
\EEQ
The homogeneous (h) differential equation then takes the form
\BEQ 
\label{g0homeqn}
  s_i^2p(p \pplus 1)  g_\hom- (\bar \Se  ^2 g_\hom')'   \pplus    \left( \frac{2}{3x}- \frac{x^2f_q}{3F_q}\right)\bar \Se  ^2g_\hom' =0 .
\EEQ
Near $x=0$, one solution  $g_\hom^\rmr=1+\cO(x^2)$ is regular (r) and  finite at $x=0$.
The singular (s) solution depends on the $x\to0$ behavior of $f_q(x)=c x^q$ with $F=c x^{q+3}/(q+3)$; 
it starts with a non-integer power, viz. $g_\hom^\rms=x^{(2-q)/3}[1+\cO(x)]$. 
However, this is not acceptable, since  (\ref{h012g}) would yield the divergencies $h_{1,2}\sim x^{-(4+q)/3}$. 
For $x\to 1$ the shape $\bar \Se  \approx s_i(x-1)$ leads to singular behaviors $g_\hom=c_1(1-x)^p+c_2(1-x)^{-p-1}$,
which leads to perturbations $h_{0,1,2}$ of this form. The $c_2$ term diverges at $x=1$, rendering it unacceptable.
The $c_1$ term is acceptable provided that it is bounded, that is, for $p\ge 0$ and $\om^2\ge 0$ through (\ref{om2p}).
So, we can already see that these perturbation modes are {\it not} unstable.

Now the tools have been collected and the conditions inspected for solving eq. (\ref{g0xeqn}) for the function $\gn(x)$.
Near $x=0$, series expansions can be performed for the inhomogeneous solution $g_\inh$ and the homogeneous
 $g_\hom^\reg$, which yield initial values at some $x_i\ll1$ for integration up to a midpoint $x_\mid\sim \half$. 
 In this regime the full solution involves an undetermined constant $a$,
\BEQ
g(x)=g_\inh(x)+a g_\hom^\reg(x) .
\EEQ
For $x$ near 1, we set $y=1-x$, define $\bg(y)=g(x)$ and set likewise
\BEQ
\gn(x)=\bg_\inh(y)+b \bg_\hom(y)=\bg_\inh(y)+b (y)^ph(y),\qquad y=1-x.
\EEQ
Series expansions  for $\bg_{\inh}$ and $h$ at small $y$ yield initial conditions at some $y_i$ for integration up to $y_\mid=1-x_\mid$.
The parameters $a$ and $b$ are fixed by matching the expressions for $g$ and $g'$ at $x_\mid$,  which works for every $p\ge0$. 
Hence, the fluctuations have a {\it continuous spectrum} for $\om$ ranging from $0$ to $\infty$.

\subsection{Connection to a Schr\"odinger problem}

The homogeneous eq. (\refl{g0homeqn}) can be expressed as a radial Schr\"odinger equation.
To show this, we define a new radial variable  $\xi$ by
\BEQ
\d\xi=-\frac{\d x}{\bar \Se  (x)} ,\qquad \xi(x)=x+\int_0^x\d x\frac{\Se  (x)}{1-\Se  (x)}  .
\EEQ
For small $x$, $\xi\approx x+\frac{2}{3}\Se  ''(0)x^3$, while $\xi\approx -(1/s_i)\log(1-x)+\text{const.}$ for $x\to1$,
so that $\xi\to\infty$ at the inner horizon. Next, we go to the function $\psi(\xi)$ set by
\BEQ \hspace{-6mm} \label{psidef}
g_\hom^\reg(x)=\frac{x^{1/3}\xi(x)\psi(\xi(x)) }{ \vert \bar \Se  (x) \vert^{1/2}F_q^{1/6}(x)} , \hspace{3mm}
\psi(\xi)=\frac{F_q^{1/6}(x(\xi))}{\xi\,[x(\xi)]^{1/3}} \vert\bar \Se  (x(\xi))\vert^{1/2} g_\hom^\reg(x(\xi)) ,
\EEQ
where $x(\xi)$ is the inverse of $\xi(x)$.
This leads to a Schr\"odinger equation with Hamiltonian operator $\hat H$ and eigenvalue $\cE$,
\BEQ \label{Schreqn}
\hspace{-6mm}
\hat H=-\frac{\d^2}{\d\xi^2}-\frac{2}{\xi}\frac{\d}{\d\xi}+V(\xi), 
\qquad
\hat H\psi=\cE\psi,\quad \cE=-\ombar ^2 =-s_i^2p(p+1).
\EEQ
The potential $V(\xi)\equiv U(x(\xi))$ is determined by
\BEQ \hspace{-6mm}
U(x)  = \frac{\Se  '^2}{4}
 \pplus  \left(\frac{\Se  ''}{2}
\pplus\frac{\Se  ' F_q'}{3 F_q}  \mmin \frac{2\Se  '}{3 x} \right)\! \bar \Se  
 \pplus 
\left( \frac{F_q''}{6 F_q} \mmin \frac{F_q'}{9 x F_q} \mmin \frac{5F_q'^2}{36 F_q^2} \pplus \frac{4 }{9 x^2}\right) \! \bar \Se  ^2 .
 \EEQ
For $f_q\sim x^q$ near $x=0$ it has a $1/x^2$ behavior, while it remains finite at $x=1$,
\BEQ
U(x)=\frac{(q+1)(q-5)}{36x^2}+\cO(x^0),\qquad V(\infty)=U(1)= \frac{1}{4}s_i^2.
\EEQ

Eq. (\ref{Schreqn}) shows that our interest lies in the bound states of $V$ with eigenvalues $\cE=-\ombar ^2 \le 0<V(\infty)$.
The leading exponential fall off  $\psi\sim \exp[-s_i(p+\half)\xi]$ corresponds via (\ref{psidef}) with $\gn\sim (1-x)^p$, as it should.

\begin{figure}\label{figrhozpRiRe} 
\centerline{ \includegraphics[width=7cm]{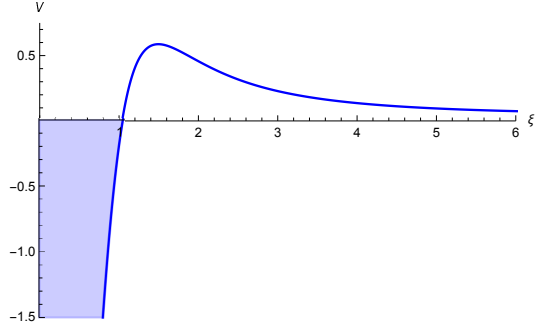}}
\caption{
The potential $V$ behaves as $-1/\xi^2$ at small $\xi$ and is positive at 
the inner horizon ($\xi\to\infty$). Fluctuations in the black hole
connect to a continuum of  bound states with $\cE<0$.
}  \label{figVxi}
\end{figure}

For the $q=4$ case of eqs. (\ref{fq-casec}), (\ref{S1-casec}) the potential $V(\xi)$ is plotted in fig. \ref{figVxi}.
The result for $U(x)$ is too long to present here.
Its small $x$ behavior reads
\BEQ
U(x)=-\frac{5}{36 x^2}
-\frac{40416715}{5657472} + \frac{5577644685915769}{352076883738624}x^2+\cdots
\EEQ
That the eigenvalues are real, results from the standard relation $I=0$, where
\BEQ\label{EmE}
I\equiv \int_0^\infty \d\xi \,\xi^2 [\psi^\ast (\hat H\psi)-(\hat H \psi^\ast)\psi]=(\cE-\cE^\ast)\int_0^\infty \d\xi  \,\psi^\ast\psi .
\EEQ
While this is correct as it is written,   for the separate contributions the small $\xi$ behavior has to be inspected.
In definition (\ref{psidef}) the finiteness of  $g_\hom^\reg$ at $x=0$ implies for $f_q\sim x^q$ that $\psi\sim \xi^{(q-5)/6}$ for small $\xi$, implying that the integrals 
\BEQ \label{q>2}
 \int^\xi \d\xi\,\xi^2\psi^\ast(\psi''+\frac{2}{\xi}\psi')\sim   \int^\xi \d\xi\,\xi^2V\vert\psi\vert^2\sim \xi^{(q-2)/3} ,
\EEQ
converge for $\xi\to0$ only when $q>2$.
For BHs with $q\le2$, inserting the Schr\"odinger equation in the right hand side of (\ref{EmE})
leads to convergent integrals, which can be combined as in the left hand side. The potential $V$ drops out and
integration yields $I=\xi^2(\psi^\ast\psi'-\psi^\ast{}'\psi)\vert_0^\infty$.  
At large $\xi$, $\psi$ vanishes exponentially.
The small-$\xi$  behavior $\psi= \xi^{(q-5)/6}(1+c\xi^2+\cdots )$ lets the dominant divergency 
$\xi^{(q-2)/3}$  cancel, leaving $\sim \xi^{(q+4)/3} $  which vanishes for all $q>-4$.

In conclusion, for any regular BH with charge density behaving as $\sim r^q$ for $r\to0$ with $q>-3$ (faster divergence would lead to infinite charge), 
the eigenmodes have real $\om^2$,
which has to be positive for relating to bounded fluctuations with parameter $p>0$. Such modes display {\it temporal oscillation but no exponential growth}.

\subsection{An explicit type of fluctuations}

Eq. (\ref{g0xeqn})  embodies a differential operator  acting on the nucleus $g(x)$, set by the unperturbed BH solution.
The source $G$ in (\ref{sourceG}) is caused by a type of fluctuations around that solution, coded by the fluctuation profile $\varrhoq(x)$. 

We apply this to  the exact solutions of section \ref{exactsols}, specified by $\Sn =1$ and a charge density profile 
set by $f_q(x)$, which determines the metric function $\bar \Se  $. 
As an example, we consider the $q=2$ case $f_q$  of eq. (\ref{fq-caseb}), with $\Se  $ in (\ref{S1-caseb}). 
For the profile $\varrhoq$ introduced in  (\ref{varrho}) and  leading to $\fqrho$ via (\refl{tildefq}),  we take
\BEQ \label{varrho4}
 \varrhoq(x) = \frac{3125}{108} x^4 (1 - x^2)^3,\qquad  
   \EEQ
 with maximum 1 for $x=\sqrt{2/5}$. With $ f_q(x)=({315}/{4})x^2(1-x^2)^2$ this leads to
 \BEQ
 \fqrho(x)= \frac{3125}{864} x^6 (1 - x^2)^2  (567 - 1935 x^2 + 1985 x^4 - 665 x^6) .
\EEQ
It holds that $\varrhoq \sim x^2f_q$ for $x\to0$, while they vanish fast enough for $x\to1$, 
properties motivated below (\ref{q>2}) and (\ref{sourceG}).  

The inhomogeneous (i) equation with $G\approx G_0$ can be solved by expansion in powers of $x^2$. Taking $g_\inh(0)=0$, the first terms are, 
with $\om_p=\sqrt{p(p+1)}$, 
\BEQ \hspace{-6mm} \label{ginhx}
g_\inh=
-\frac{114003125}{4080845772}  \om_p^2 x^6- 114003125\frac{ 2119516335+1021468 \om_p^2}{7675596082564736256}\om_p^2x^8.
\EEQ
The next term is already much longer and would not fit in one line. 

The regular (r) solutions starts as
\BEQ\hspace{-6mm} \label{hhomx} \label{ghomx}
g_\hom^\rmr = 1+\frac{36481}{4198401}\om_p^2x^2+
\frac{1820481177+2042936\om_p^2 }{3948351894323424}\om_p^2x^4.
\EEQ

Because the singular power $x^{(q-2)/3}$ collides with $x^0$ in this $q=2$ situation, 
the singular homogeneous solution has the form of $g_\hom^\rms(x)=g_\hom^\rmr(x)  \log x +\tilde g_\hom^\rms(x)$, 
where $\tilde g_\hom^\rms$ has a series expansion in powers of $x^2$, starting as 
\BEQ \hspace{-9mm}
1 \pplus  \frac{314371}{152992 }x^2 \pplus \!\left( \! \frac{7520530571161}{2527907622912} \pplus \frac{32412383513 \om_p^2}{3853930594752} \mmin
\frac{1330863361 \om_p^4}{141012567654408} \! \right)x^4 \! .
\EEQ 
As pointed out, the singular solution has to be discarded.

The behavior near $x=1$ is complicated but provides interesting insights. 
For $\bg (y)\equiv \gn (\sqrt{1-y})$, a series expansion is possible in powers of $y=1-x^2$.
(In absence of odd powers of $x$, this definition is more economic than $y=1-x$.) 
The inhomogeneous (i) solution $\bg_\inh(y)$ of  (\ref{g0xeqn}), (\ref{sourceG}) has the leading terms
\BEQ \label{ginhy}
&&
\bg_\inh=
\frac{3125 p (1 + p) }{108 (p-3) (p+4)} y^3 -\frac{3125 p (1 + p) (764 p(p+1) -49453  ) }{
 41256 (p-4 ) (p-3) (p+4) (p+5 ) } y^4\qquad
\EEQ
For a general source expandable in powers of $y$ and starting as $y^k$, $\bg_\inh$ has  the schematic pole structure 
$\sum_{l=0}^\infty y^{k+l}/[(\om_p^2-\om_k^2)\cdots (\om_p^2-\om_{k+l}^2)]$. 
As seen from (\ref{varrho4}) and (\ref{sourceG}), the present source $G_0$ has $k=3$,  while $G_1$ has $k=1$ and $G_2$ has $k=2$.
This pole structure can be understood as follows. The leading small-$y$ behavior of (\ref{g0xeqn}) is $s_i^2[\om_p^2\bg_\inh-2ybg_\inh'-y^2\bg_\inh'']=G$.
For $G=g_ky^k$ the leading term is  $\bg_\inh=g_k y^k/s_i^2(p-k)(p+k+1)$.
The mixing in of this pole in the higher order coefficients comes from acting as a source for them.

As discussed below (\ref{g0homeqn}), the homogeneous (h) solution has leading behavior $y^p$.
In the present situation it takes the form $\bg_\hom(y)=y^p \bh(y)$ with
\BEQ \label{hhomy} \hspace{-8mm}
\bh=1 -\frac{7866 p+16687}{2292 (p+1)} py +\frac{61873956 p^2+276084648 p+247575415}{5253264 (2 p+3)}py^2 .
\EEQ
By going to 25'th order, it is seen that the $k$'th order coefficients in $\bg_\inh$ and $\bh_p$ diverge roughly as $1/y_0^k$ with $y_0\sim 0.02$.
Hence the series converge at $y<y_0$, where they can be used for the initial values in a numerical integration.

Interestingly,  the ratio $g_\inh/h$ has the simpler single-pole structure,
\BEQ \hspace{-3mm}
\frac{g_\inh(y)}{h(y)} \iss \frac{3125 p (p+1) y^3}{108 (p \mmin 3)(p \pplus 4)} \pplus
 3125 p\frac{ 7866 p^3 \pplus 43567 p^2 \mmin 19100 p \mmin 98906 }{247536(p-4) (p+4) (p+5)}y^4,
\EEQ
which has the general pole structure of schematic form $\sum_{l=0}^\infty y^{k+l}/(p-k-l)$, here with $k=3$.
This follows analytically by inserting $\gn_\inh=g_\hom j_\inh=(y^pj_\inh)h$ in  (\ref{g0xeqn}), which eliminates $\ombar ^2j_\inh$ and leads to 
$j_\inh'\sim\sum_k g_k y^{k-p-1}/s_i^2(p+k+1)$.

The full solution for $0\le x<x_f$ combines the (in)homogeneous solutions, 
\BEQ
g(x)=g_\inh(x)+a g_\hom(x)
\EEQ
for some $a$. For $x$ between $x_f$ and 1 the results for $0<y<y_f$ yield
\BEQ
\gn(x)=\bg_\inh(y)+b \bg_\hom(y) =\bg_\inh(y)+b \, y^ph(y) , \qquad y=1-x^2.
\EEQ
The coefficients $a$ and $b$ are determined by matching the expressions for $\gn$ and $\gn'$ at a midpoint $x_m\sim\half$, $y_m=1-x_m^2$.
Since the (in)homogeneous solutions exist for all $p$, so does the full solution $g$. All positive values of $p$ being allowed means that the 
frequency spectrum for this type of fluctuations is a continuum, running from $\om=0$ to $\infty$.

\begin{figure}\label{fig-g-var-p.eps} 
\centerline{ \includegraphics[width=6cm]{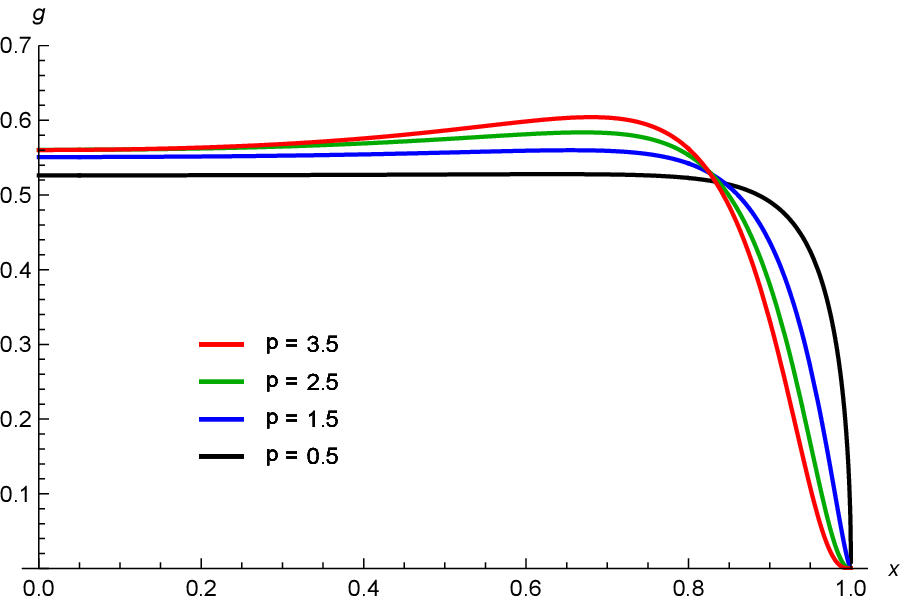} \, \includegraphics[width=6cm]{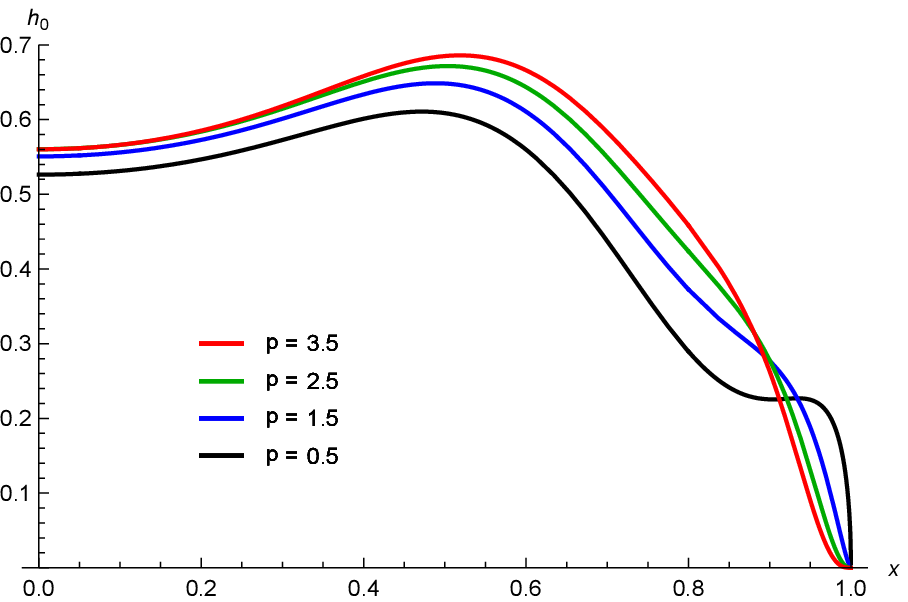}}
\caption{
The nucleus $g(x)$ and the metric perturbation $h_0(x)$, plotted for various spectral parameters $p$, 
are finite at the  origin $x=0$ and vanish at the inner horizon $x=1$.
\label{figg0h0}}
\end{figure}

\begin{figure}\label{fig-g-var-p.eps} 
\centerline{ \includegraphics[width=6cm]{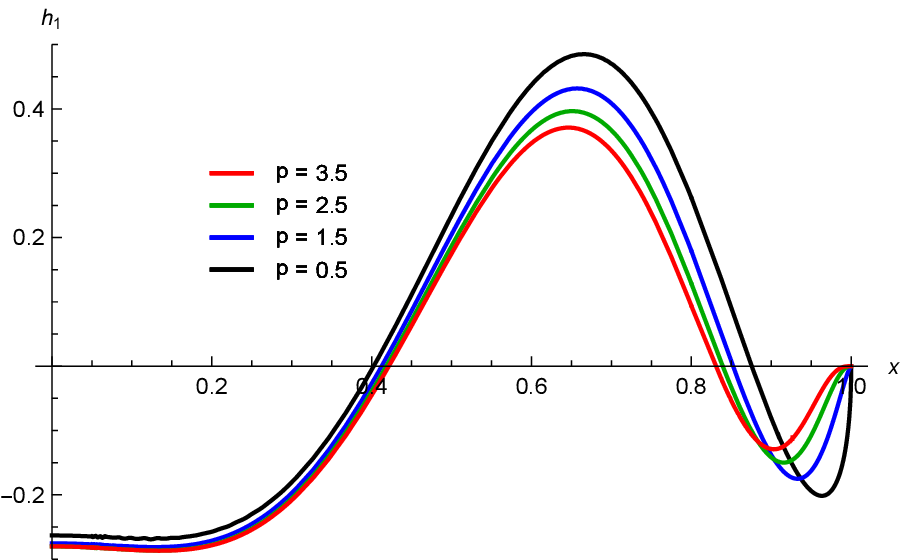}  \includegraphics[width=6cm]{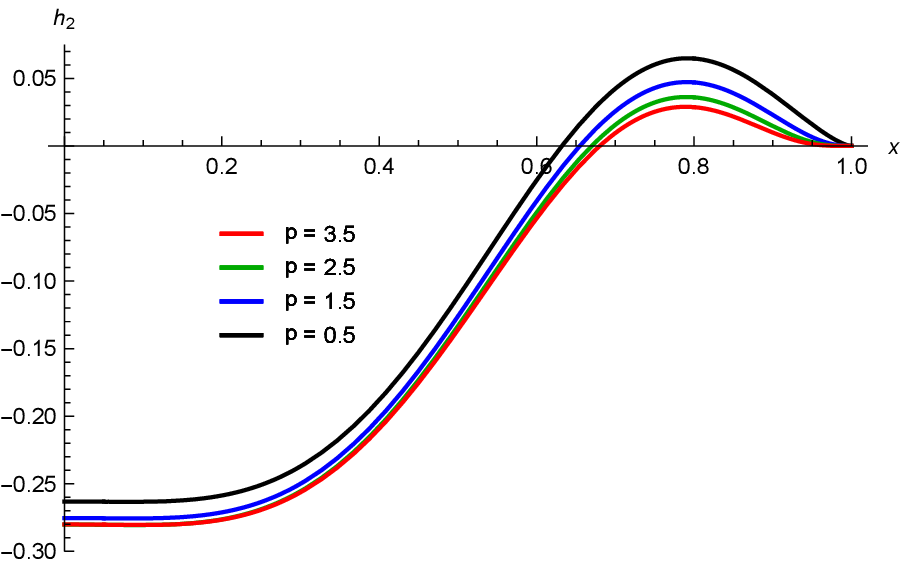}}
\caption{
The metric perturbations $h_{1,2}(x)$, plotted for various spectral parameters $p$, 
are finite at the origin $x=0$ and vanish at the inner horizon $x=1$.}
\label{figh1h2}
\end{figure}

In figures \ref{figg0h0} and \ref{figh1h2} we plot the nucleus $g$ and the metric functions $h_{0,1,2}$ as a function of $x$.
They vanish at the inner horizon, as they should, and are otherwise bounded. The possiblity to construct them at the chosen values of $p$ reflects
the presence of the continuum spectrum of fluctuation modes. 

\renewcommand{\thesection}{\arabic{section}}
\section{Additional boundary shells}
\setcounter{equation}{0}
\renewcommand{\theequation}{10.\arabic{equation}}

\label{boundaries}

The above solution works for large charge--to--mass ratios of the BH. 
Extension to other values is explored now.

\subsection{Charges on the outside of the inner horizon}

To start, let us recall that in the mantle,
a decreasing radial parameter $r$ plays the role of increasing time, so that particle trajectories progress towards the inner horizon.  
We consider a  charge current in a thin layer beyond $R_i$, for $x\equiv r/R_i$ in the range $1\le x\le x_l$, 
and take the limit $x_l\downarrow1$ at the end.

We assume that the core is still defined by the dimensionless charge distribution $f_q(x)$ with its $I_q$ set by (\ref{si-Iq}), 
while the inner horizon $R_i$ is allowed to take a modified value.
The core has mass total $\Mc $ and total charge $\Qc $, and  the mantle total mass $M_m$ and total charge $Q_m$.

In the mantle, the charge current reads $\vJ=(J^0,J^2,J^3)$ and the vector potential $\vA=(A_0,A_2,A_3)$. 
The electromagnetic potential $A_\mu=\delta^0_\mu A_0(r)$ implies ${\bf B}={\bf 0}$ and $\vE=(E,0,0)$ with $E=-A_0'/\Sn $.
 The Maxwell law $\nabla\times\vB-\dot\vE=\mu_0\vJ$ reads in the BH mantle $(E'+2E/r)/\Sn =\mu_0J^0=\mu_0\rho_q/\Sn $, 
 the very relation of section \ref{electro} for implementation of $\nabla\cdot\vE=\mu_0J^0$ in the core.

Let us keep $\Sn =1$ and consider in the mantle region $1\le x\le x_l$ the shape
\BEQ \label{S1exactgeneralm} \hspace{-3mm}
\Se  (xR_i) =\frac{2GM_m}{xR_i}-\frac{GQ_m^2}{x^2R_i^2}-\frac{4GQ_m^2}{3R_i^2}
\int_{x}^{x_l} \d y\left ( \frac{x^2}{y^3}-\frac{1}{x}\right)\frac{\Phi(y)} {y^2}  .
\EEQ
The first two terms represent a RN metric. 
The function $\Phi\ge 0$ is decaying, $\Phi'\le 0$ for $1\le x\le x_l$, while $\Phi=0$ for $x\ge x_l$.
When taking the limit $x_l \downarrow 1$, we assume some finite value for $\Phi(1)$.
The integral in (\ref{S1exactgeneralm}) vanishes for $x>x_l$ and for $1\le x\le x_l$ also in the limit $x_l\downarrow 1$, 
so that $M_m$ and $Q_m$ are the mass and the charge in the mantle, respectively. Likewise, the RN value 
 for $s_i^+=R_i\Se  '(R_i^+)$ emerges for $x_l\downarrow 1$.
 In the mantle, eq. (\ref{rhoA=}) holds after changing sign of the $\Sn '\bar \Se  $ term.
Inserting Eq. (\ref{S1exactgeneralm}) and $\Sn =1$ leads to $\rhoA =Q^2(r)/8\pi r^4$ with
\BEQ
Q(r)=Q_m\sqrt{1- \Phi (x)},\quad Q_c =Q_m{\color{black}\sqrt{\Phi_q^c}},\quad \Phi_q\equiv \Phi(1)\equiv 1-\Phi_q^c,
\EEQ
evidently imposing that $\Phi_q\le 1$. Since $-Q'(r) \propto \Phi'(x) \le 0$, the situation corresponds to a current of positive charges towards the inner horizon.

\newcommand{\Phic}{\Phi_q^c}

Also in the mantle, the local cosmological constant follows from (\ref{rhoL=}); Eq. (\ref{S1exactgeneralm}) yields
\BEQ
\rho_\lambda(r)=\frac{Q_m^2}{8\pi R_i^4}\left[ \frac{\Phi(x)}{ x^4} - 4\int_x^{x_l} \d y \frac{\Phi(y)}{y^5}\right],
\EEQ
which is nonnegative and vanishes for $x\ge x_l$.
The finite value at $R_i$,
\BEQ\label{roLRi}
\rho_\lambda(R_i^+) 
\approx  \frac{Q_m^2}{8\pi R_i^4}\Phi_q=\frac{1}{8\pi GR_i^2}\frac{\qi^2\Phi_q}{\Phi_q^c} , 
\EEQ
changes the fit to the interior; equating to (\ref{rhoLR<}) yields, for the $I_q$ set by the core, 
\BEQ
 \qc ^2 =\frac{3\Phic }{1+4\Phic I_q},\qquad
 s_i^- =2-4\qi^2I_q=2\frac{1-2\Phic I_q}{1+4\Phic I_q}.
 \EEQ
For general $x_l$ the latter has to fit  $s_i^+\approx 2p_m/(1-p_m)$, where $p_m=\sqrt{1-q_m^2}$ with 
 the charge-to-mass ratio $q_m=\cQ _m/M_m$. This yields for $x_l\downarrow 1$
 \BEQ
 p_m=\frac{\,\, 1-2\Phic I_q}{2(1+\Phic  I_q)} ,\qquad
 R_i=(1-p_m)GM_m=\frac{1+4\Phic I_q }{1+\Phic  I_q}\frac{GM_m}{2} .
 \EEQ
 While the layer current coded by $\Phi_q$ generally makes $p_m$ larger and  hence $q_m$ smaller,
  the maximum $p_m=\half$ occurs for $\Phic I_q\to0$, so that still $q_m\ge\half\sqrt{3}$.

 From $\Se  (R_i^-)=GM_c/R_i-GQ_c^2/R_i^2=1$ it follows that
  \BEQ
 \frac{ M_c}{ M_m}= 1-\frac{3 \Phi_q}{  4 (1 +\Phic  I_q)}  , \quad 
  \frac{\cQ _c}{M_m}=\frac{ \sqrt{ 3\Phic (1 + 4\Phic  I_q)}} {2(1 + \Phic  I_q )} ,
  \EEQ
and
\BEQ
  \frac{\cQ _c}{M_c}=\frac{2 \sqrt{ 3\Phic (1 + 4 I_q\Phic)}} {4 + 4\Phic  I_q - 3 \Phi_q} .
 \EEQ
In the limit $\Phi_q\to1$ ($\Phi_q^c\to0$), the core charge vanishes and $M_c\to M_m/4$, as was the case
in subsection \ref{toRi-} when all charges in the core were located against the inner horizon.
When core charges exist $(\Phi_q^c>0)$, but get located towards the inner horizon, one has $I_q\to0$, so that
$M_c/M_m\to 1-3\Phi_q/4>\frac{1}{4}$ due to the shift of $\rho_\lambda$. Next,
 $\cQ _c/M_m\to\half { \sqrt{ 3\Phic }}$, while still $\cQ _m/M_m\to\half  \sqrt{ 3}$.

In the mantle, the velocity vector entering eq.  (\ref{TMnrhomA}) reads $U^\mu=\delta_1^\mu\sqrt{\bar \Se  }$, 
so that eq. (\ref{sigth=}) acquires an extra minus sign in the right hand side, consistent with $\sigma_\vth\ge0$. 
But the factor $\bar \Se  \approx s_i^\pm (x-1)\le s_i^\pm(x_l-1)\downarrow0$ is incompatible with a sizeable 
thermal mass component in the boundary layer.

This layer is locked up since it cannot go back to larger $r$ and neither pass the inner horizon at $R_i$ during the aeon of an external observer.

\subsection{Sauter-Schwinger blanket around the event horizon} 

As discussed in section \ref{estimates}, the charge of a BH can be annulled by $e,\bar e$ creation near the event horizon,
whereby the positrons escape to infinity while the electrons form a blanket around the horizon, thereby  {\it lowering} the BH mass.
Accreted matter can compensate the remaining charge.
The relation $M_m-Q_m^2/2R_e=M-Q^2/R_e$ with $Q=0$ exhibits  energy conservation:
the small mass fraction $\sim m_e /em_P$ of positive charges carry the electrostatic energy $Q_m^2/2R_e$  away to infinity,   
so this part of $M_m$ is no longer counted in the black hole energy (mass) $M$. 
In doing so, they compensate the charge of the excess electrons ejected in the stellar collapse.

\renewcommand{\thesection}{\arabic{section}}
\section{Summary} 
\setcounter{equation}{0}
\renewcommand{\theequation}{\thesection.\arabic{equation}}

A class of exact solutions for the interior of charged, nonrotating astrophysical black holes (BHs) is presented.
It is supposed that in the BH formation by stellar core collapse,  the nucleons reach a high density and dissolve in up and down quarks.
The released binding energy can lead to an energy density at the weak scale, $\sim\lambda v^4$, 
which applies to objects with  at least 1.5 Neptune mass; in practice:  to solar masses and beyond. 

The solutions have an event horizon and an inner horizon, but no singularity.
Matter is located in the core bounded by the inner horizon; outside it, there is 
a mantle having a standard vacuum with Reissner Nordstr\"om metric.  
This setup works owing to the negative pressure of the zero point energy and the
negative radial pressure caused by the electric charges.
Additional charge layers may be present on the outside of the inner and event horizons.
Hence, an extremal BH interior can be neutral.

Spherically symmetric perturbations appear to have a continuous oscillation spectrum, without modes growing in time.
With the core having a normal time coordinate and regular properties, this shows that our BHs are also well behaved at the level of fluctuations.
The absence of a discrete frequency spectrum is likely connected with the mild conditions we put. 

Our novel solutions bear on the capability of the quantum fields to absorb some 25\% of the binding energy in their zero point energy,
the BH equivalent of the Casimir effect. For the cases (\ref{goodQr}) and (\ref{secondfq}) the related profiles are plotted.
The zero point energy density is  maximal in the center and vanishes smoothly at the inner horizon.
 
 In a first step, the rest energy density and pressure of the collapsed matter is neglected and zero matter temperature is assumed.
 This leads to an analytical solution given the charge density profile; several examples are considered. Incorporating the collapsed fermions
  leads to a numerical approach yielding corrections at or below the per cent level.
A general approach would allow a finite temperature profile $T(r)$, whereby a Higgs condensate and pairs of the standard model (anti)particles exist,
to be treated in renormalized thermal quantum field theory. This problem can be simplified significantly by approximating all particles as massless.

\renewcommand{\thesection}{\arabic{section}}
\section{Discussion}
\setcounter{equation}{0}
\renewcommand{\theequation}{\thesection.\arabic{equation}}

The problem of regularizing the stress energy tensor in general relativity is celebrated, and even studied recently \cite{taylor2022mode}. 
Point splitting \cite{christensen1976vacuum} leads to a finite, cumbersome result with extra geometric terms. 
These are of order $\delta\rho\sim m_P^8/M^4$, very much smaller than our typical scale $\lambda v^4$, 
so this issue has no bearing on our case.

It is known that BHs can not be over-charged or over-spun \cite{wald2018kerr}. 
 BHs can not be made over-extremal  (in charge, combined with rotation) to make the horizons disappear,
at least temporarily, and make fireworks possible.  In this classical approach it is not possible to connect to the Fermi bubbles \cite{su2010giant} 
and the X-ray chimneys \cite{ponti2019x} around Sag A$^\ast$, as consisting of electrons released from the BH,
but see  \cite{blandford1977electromagnetic} for a modelling of the ergosphere.

\renewcommand{\thesection}{\arabic{section}}
\section{Outlook} 
\setcounter{equation}{0}
\renewcommand{\theequation}{\thesection.\arabic{equation}}

Our charged BH model can act as a stepping stone towards rotating BHs. We plan to present
a minimal regularization of the Kerr Newman metric elsewhere.

When the mantles are much larger than the cores, the latter remain shielded  in BH-BH merging and our theory is not tested.
For extremal BHs it is tempting to assume that the horizons will temporarily open up during merging events, that produce gravitational waves.
The merging of two extremal BHs, or perhaps an extremal BH with a normal BH, or an extremal BH with a neutron star, may expose the interior(s).
Firstly, the gravitational wave signal in a merging event may be modified by the non-trivial cores, which can be tested in high signal-to-noise events.
Secondly, the full spectrum of electromagnetic waves (``fireworks'') may be generated, akin to the multi-messenger observations 
for a binary neutron star merging \cite{hartley2017multi}.
Observation of such events for a mass exceeding the bound of $2.35M_\odot$ \cite{romani2022psr} 
support our theory, and may have occurred already in the event S190426c \cite{Ligo2019a,Ligo2019b,lattimer2019properties}. 

The present work offers a new view on astrophysical BHs and a further application of the standard model of particle physics.
Many questions remain, for example:  Can this class of metrics be reached dynamically?
Are there (meta)stable ones?  What are criteria for optimized profiles? 
What is the role of their various entropies \cite{nieuwenhuizen2007role}? Can the approach be generalized to matter at
finite temperature and to rotating BHs? It can actually be applied to the dark matter problem, see \cite{nieuwenhuizen2023solution}.

\section*{Acknowledgements}

Discussion with Iosif Bena, Eric Laenen, Tomislav Prokopec, J\'er\^ome Houdayer, Jan Smit, Roger Balian and Peter Keefe
is gratefully appreciated.


 \end{document}